\documentclass[twocolumn, secnumarabic,amssymb, nobibnotes, prl, aps,showkeys]{revtex4-2}

\usepackage{graphicx}
\usepackage{amsmath}
\usepackage{float}
\usepackage{lipsum}

\usepackage{lineno,hyperref}
\modulolinenumbers[5]

\begin{document}

\title{Time calibration studies for the Timepix3 hybrid pixel detector in electron microscopy}

\author{Yves Auad}
 \email{yves.maia-auad@universite-paris-saclay.fr}
\author{Jassem Baaboura}
\author{Jean-Denis Blazit}
\author{Marcel Tencé}
\author{Odile Stéphan}
\author{Mathieu Kociak}
\author{Luiz H. G. Tizei}
\affiliation{%
Laboratoire des Physique des Solides, Université Paris Saclay, Orsay, France
}%
\date{\today}


\begin{abstract}
Direct electron detection is currently revolutionizing many fields of electron microscopy due to its lower noise, its reduced point-spread function, and its increased quantum efficiency. More specifically to this work, Timepix3 is a hybrid-pixel direct electron detector capable of outputting temporal information of individual hits in its pixel array. Its architecture results in a data-driven detector, also called event-based, in which individual hits trigger the data off the chip for readout as fast as possible. The presence of a pixel threshold value results in an almost readout-noise-free detector while also defining the hit time of arrival and the time the signal stays over the pixel threshold. In this work, we have performed various experiments to calibrate and correct the Timepix3 temporal information, specifically in the context of electron microscopy. These include the energy calibration, and the time-walk and pixel delay corrections, reaching an average temporal resolution throughout the entire pixel matrix of $1.37 \pm 0.04$ ns. Additionally, we have also studied cosmic rays tracks to characterize the charge dynamics along the volume of the sensor layer, allowing us to estimate the limits of the detector's temporal response depending on different bias voltages, sensor thickness, and the electron beam ionization volume. We have estimated the uncertainty due to the ionization volume ranging from about 0.8 ns for 60 keV electrons to 8.8 ns for 300 keV electrons. 
\end{abstract}



\maketitle


\section*{Introduction}

In recent years, scanning transmission electron microscopy (STEM) has been profoundly transformed by the improvements of multiple technologies, such as aberration correction and electron monochromators. Electron detection followed the revolution, mostly by the advent of direct electron detectors, providing a reduced point-spread-function and an increased quantum efficiency relative to their predecessors that used a scintillator layer. Today, the superiority of direct electron detectors is indisputable, confirmed by the extensive and fast-growing number of results concerning imaging \cite{van2020sub}, 4D STEM \cite{jannis2022event}, and electron energy loss spectroscopy (EELS) \cite{hart2017direct, tence2020electron, auad2022event}.

\begin{figure*}[t!]
    \centering
    \includegraphics[width=0.8\textwidth]{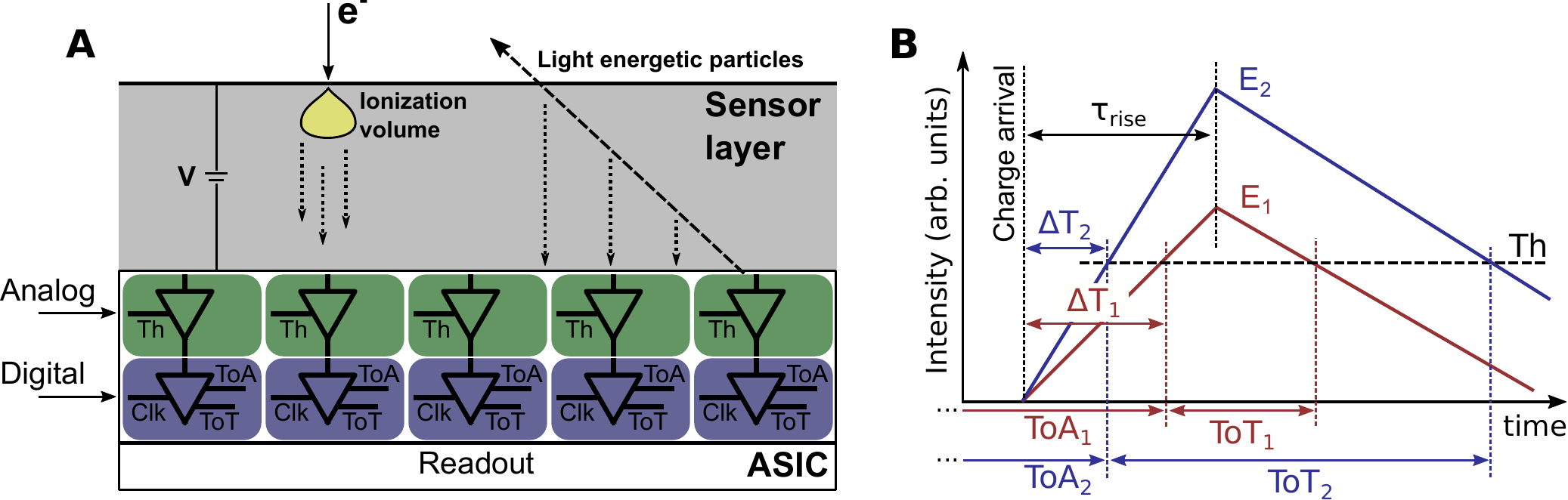}
    \caption{\textbf{General schematic of the detector and the different measured quantities.} (\textbf{A}) The TPX3 HPD consists of a sensor layer and an ASIC. In an electron microscope, charges are created at one side of the detector, which moves towards the ASIC side upon an applied bias (V). Different ionizing particles produce distinct mechanisms of charge creation. As an example, a light energetic particle, such as a muon, traverses the detector and creates charges throughout the sensor layer thickness. The charges collected at the ASIC is detected by an analog circuit depending on the pixel threshold (Th), and are timestamped by the digital part of the circuit, in particular, the ToA and ToT, using the distributed clock (Clk) signal through the pixels. (\textbf{B}) Schematic of the ToA and the ToT. The roughly constant rise time $\tau_{rise}$ of the impulsion in the analog circuit produces a ToA value dependent on the signal intensity (or, equivalently, on the ToT or the total energy (E) deposited by the hit). This effect, known as time-walk, produces major discrepancies between the received ToA and the actual electron arrival time.}
    \label{Figure1}
\end{figure*}

One kind of direct electron detector is the so-called hybrid pixel detector, named as this because the semiconductor sensor layer and the application-specific integrated circuit (ASIC) are independently manufactured \cite{ballabriga2018asic}. For the concern of this paper, the Timepix3 (TPX3) is an event-based detector, capable of outputting temporal and positional information of individual electron hits. Each pixel possesses its individual electronics, comprising an analog and a digital processing circuitry \cite{poikela2014timepix3}. A threshold value defines the minimal input signal intensity the pulse must have to be considered a pixel hit, and it can be set, pixel-by-pixel, on the analog processing part of the pixel electronics, allowing a virtually complete suppression of the readout noise of the detector. The temporal information of the pixel hit is given by the instant the analog signal surpasses the pixel threshold value, called time of arrival (ToA), and the time duration the analog signal is kept over the pixel threshold value, called time over threshold (ToT). The ToA and ToT are after products of the digital processing part, and are latched on the distributed clock in the pixel array, as can be seen in Figure \ref{Figure1}. While the ToT reaches a time bin of 25 ns from the 40 MHz clock frequency, the ToA is further refined by a 640 MHz voltage-controlled oscillator, reaching thus a 1.5625 ns time bin. Additionally, panel \ref{Figure1}B exemplifies how the ToA value obtained is longer than the actual charge arrival time, which can be properly corrected with the combined knowledge of both ToA and ToT, as discussed later. These properties have recently enabled readout-free, live-processing EELS data reconstruction at the speed of typical imaging detectors ($\sim$ 40 ns per pixel in our case) by synchronizing the scanning unit and the TPX3 clocks \cite{auad2022event}. Such technology makes possible nanosecond-resolved temporal resolution in EELS, but can also provide a robust solution for sensitive samples, in which custom scan patterns have been suggested to help \cite{stevens2018sub, zobelli2020spatial}. Additionally, Timepix3 has also enabled the performance of the so-called cathodoluminescence excitation spectroscopy (CLE), in which the temporal correlation of electrons and infrared/visible/ultraviolet photon pairs can circumvent the absence of resonant experiments with fast electrons due to their broadband excitation spectra \cite{varkentina2022cathodoluminescence, feist2022cavity}. These techniques can be combined together, providing hyperspectral imaging of correlated electrons and thus the spatial information of the excitation pathways \cite{varkentina2022cathodoluminescence}. Although coincidence experiments can also be performed with X-rays photons, x-rays detectors have a poor temporal response, typically two orders of magnitude higher than the minimal bin of Timepix3. For visible-range photons, as in CLE, on the contrary, photon counting with photomultiplier tubes can reach sub-nanosecond temporal resolution, which gives access to the dynamics of the process in the range of the TPX3 time bin \cite{varkentina2023excitations}. Pushing the temporal resolution of TPX3 can also be interesting for performing electron energy-gain spectroscopy (EEGS) \cite{de2008electron} in continuous-gun electron microscopes, in which typical approaches rely on the usage of electrostatic beam blankers, and, in some cases, high voltages are needed, undermining the design of high-repetition rates switching circuits \cite{das2019stimulated, auad2023muev}. With proper-calibrated TPX3, repetition rates of tens of MHz should be possible, and energy-gain experiments can be performed very similarly as in CLE, with the distinction that pairs are between the injected photons and the inelastically scattered electrons.

Unfortunately, approaching the nominal TPX3 temporal resolution uniformly throughout the entire pixel matrix is not straightforward \cite{jakubek2011precise, turecek2016usb, bergmann20173d, pitters2019time, wen2022optimization}. It requires a good understanding of both parts of HPDs. The fast electrons impinging in the silicon sensor create electron-hole pairs that will drift towards the opposite side of the layer due to an applied bias. For fast electrons typically within the 30 - 200 keV energy range, theses charges are often collected by distinct pixels, creating thus clusters: multiple hits originated by the same incident electron. This process can be readily identified during data processing by spatially and temporally comparing pixel hits. The drift time depends on the electric field profile inside the silicon slab, and hence on the voltage bias applied. Additionally, the charges are created in a spatial profile that depends on the electron energy (the so-called "ionization pear" or ionization volume model), which consequently can result in slightly different charge collection intervals. Upon arrival in the individual pixel readout electronics, in the ASIC part of the detector, the digital conversion of this time of arrival can reach the aforementioned nominal value of 1.5625 ns. Understanding all these steps, from the impinging electron to the digital conversion of the charge time of arrival is important to reach the detector's best possible temporal response. 

In particular, one of the major time calibration steps is the correction of the time-walk effect, a consequence of the roughly constant rise time $\tau_{rise}$ of the analog part of the ASIC circuit and that produces a temporal shift $\Delta T$ between the latched ToA and the actual charge arrival time in the ASIC, as illustrated in Figure \ref{Figure1}B. Comparing the orthogonal triangles with heights $Th$ (pixel threshold) and $E$ (pixel deposited energy), the expected time interval is $\Delta T = \tau_{rise} Th / E$. In a more generalized way, time-walk can be modeled with the following equation:

\begin{equation}
    \Delta T(x,y, E) = \frac{a(x, y)}{E-b(x, y)} + c(x, y)
    \label{Equation1}
\end{equation}

\noindent
where $a, b, c$ are the constants that must be determined, and $x$ and $y$ are the pixel coordinates. Finally, the distributed clock net along the pixel array is imperfect and not instantaneous; thus, spatially-dependent pixel relative times can also happen. The temporal resolution of the detector, roughly speaking, is thus the propagated uncertainties of both the uncertainty associated with the time-walk correction and the uncertainty related to the time delay estimate. For the former, the contribution is multi-factorial: it depends on the discriminator jitter (a temporal uncertainty linked to when the signal went over the established threshold), on the bin size of the fine ToA, and, particularly for our study, the uncertainty related to the ionization volume of the electron beam.

To the best of our knowledge, there are no complete TPX3 temporal calibration studies in the context of electron microscopy, and all aforementioned time calibration works are performed with X-rays photons or highly energetic particles ($>$ MeV). In this work, we present a methodological study of the impact of the temporal calibration of the Timepix3 detector for electron microscopy using fast electrons (20 - 100 keV) as the source of the charge creation in the sensor layer. Besides, we stick with calibration procedures that primarily rely on data/cluster analyses from an electron beam illumination dataset without using more intricate methods, such as test pulsing calibration \cite{pitters2019time}, that, although very precise, requires more hardware manipulation. We begin by analyzing a method for the energy calibration of the detector, i.e., the relation between the ToT and the deposited energy. Next, the time-walk is corrected using a flat-field electron illumination dataset. Finally, electron-photon pairs are used to compensate for the non-uniform clock distribution net and also for verifying the calibration after the aforementioned steps. In the conclusion, we attempt to estimate the ultimate temporal resolution for the Timepix family of detectors in electron microscopy by analyzing cosmic rays tracks of light energetic particles and relating them with the ionization volume of fast electrons. 

\begin{figure}[t!]
    \centering
    \includegraphics[width=0.32\textwidth]{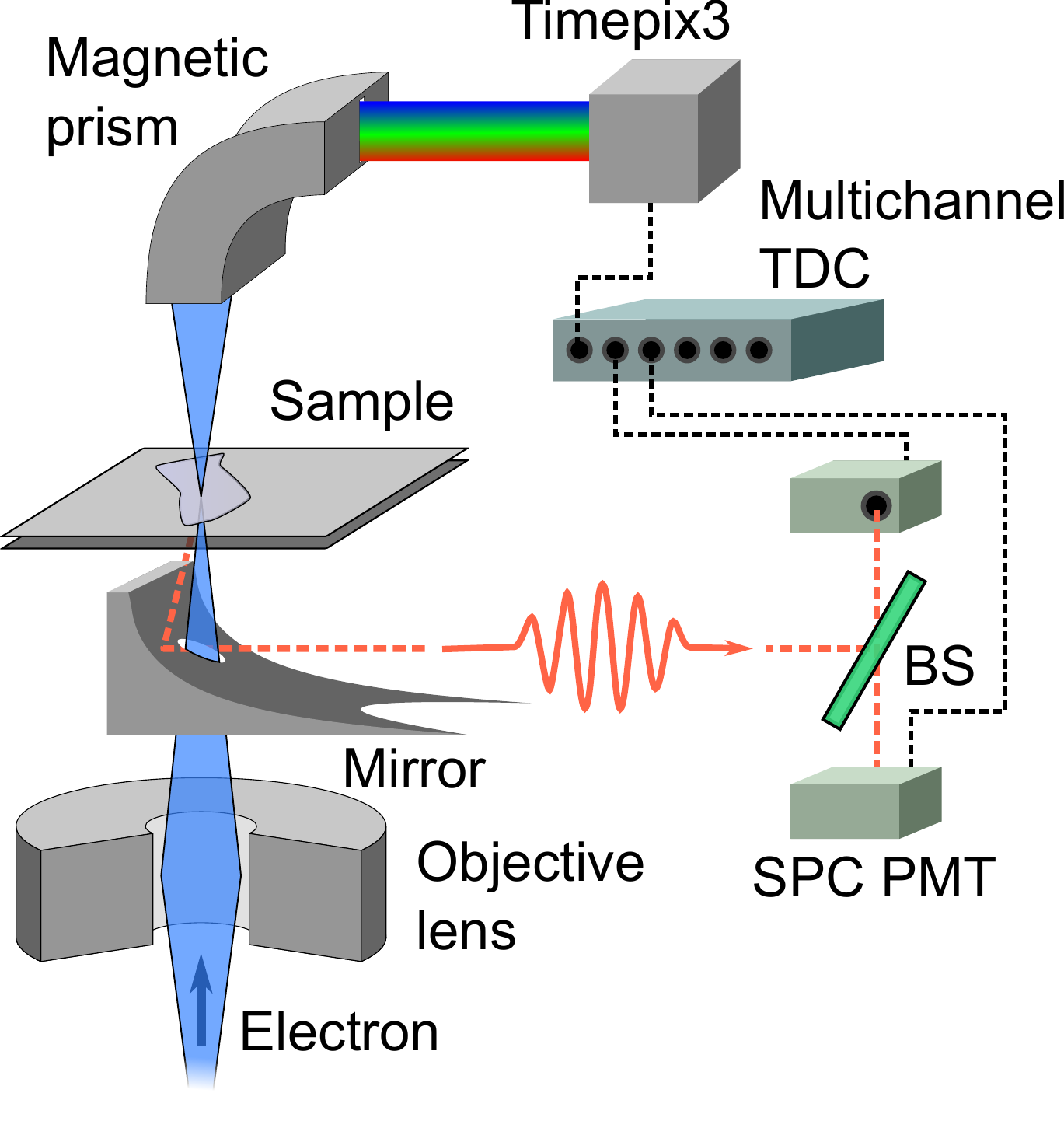}
    \caption{\textbf{Schematic of the experiment set-up}. The fast electron (20 - 100 keV) is transmitted through the sample and the electron beam is sent to TPX3 after being deflected by an electron spectrometer (magnetic section). The emitted light is collected by a parabolic reflector coupled to an optical fiber. Light can be sent to one or two single-photon counting photomultiplier tubes. A multichannel time-to-digital converter unit allows us to temporally correlate electrons and photon events. SPC-PMT: single-photon counting photomultiplier tube. BS: beamsplitter. TDC: time-to-digital converter.}
    \label{FigureScheme}
\end{figure}

A scheme of the complete experimental setup is shown in Figure \ref{FigureScheme}. A finely focused electron beam between 20 keV - 100 keV in energy is transmitted through the sample, and reaches a magnetic sector which disperses the electron beam in energy in the Timepix3 detector, a technique called electron energy-loss spectroscopy (EELS). Upon crossing the sample, light may be emitted, a process known as cathodoluminescence (CL), and photons can be guided either to a unique single-photon-counting photomultiplier tube (PMT), either to a beamsplitter and thus to two PMTs, in which Hanbury-Brown-Twiss (HBT) interferometry can be performed. With HBT, photon bunching processes \cite{meuret2015photon} can be used to extract the optical excitation's lifetime with a better temporal resolution than TPX3, thus providing a benchmark for the measured value. A home-made multichannel time-to-digital converter (TDC) with a temporal bin of 120 ps is interfaced between the Timepix3 and the two PMTs, allowing to compare all these sources of events temporally. For this work, we have used the Timepix3 commercialized by Amsterdam Scientific Instruments (ASI) called CheeTah, an array of 4 chips disposed as 1024 x 256 pixels. The detector is mounted in a Vacuum Generators HB501, a STEM dedicated microscope with a cold field emission gun and a typical spectral resolution of 300 meV in EELS. The CheeTah solution also has 2 TDC inputs capable of reaching a time bin of 260 ps. Note that for some calibration steps, no sample is needed, and sometimes a single PMT may be used directly in the TPX3 TDC input. In any case, the exact experimental condition for each step is detailed in the text. Otherwise stated, curves are generally fitted with Gaussians, and the referred temporal resolution in this work is used as a synonym for the standard deviation of the fitting result. Finally, we have used our own software for cluster identification and raw data processing. The MIT-licensed open-source software is entirely coded in Rust programming language and can also be used for live data processing of several acquisition modes \cite{auad2022tp3tools}.

\section*{Results \& Discussion}

\begin{figure*}[t]
    \centering
    \includegraphics[width=0.62\textwidth]{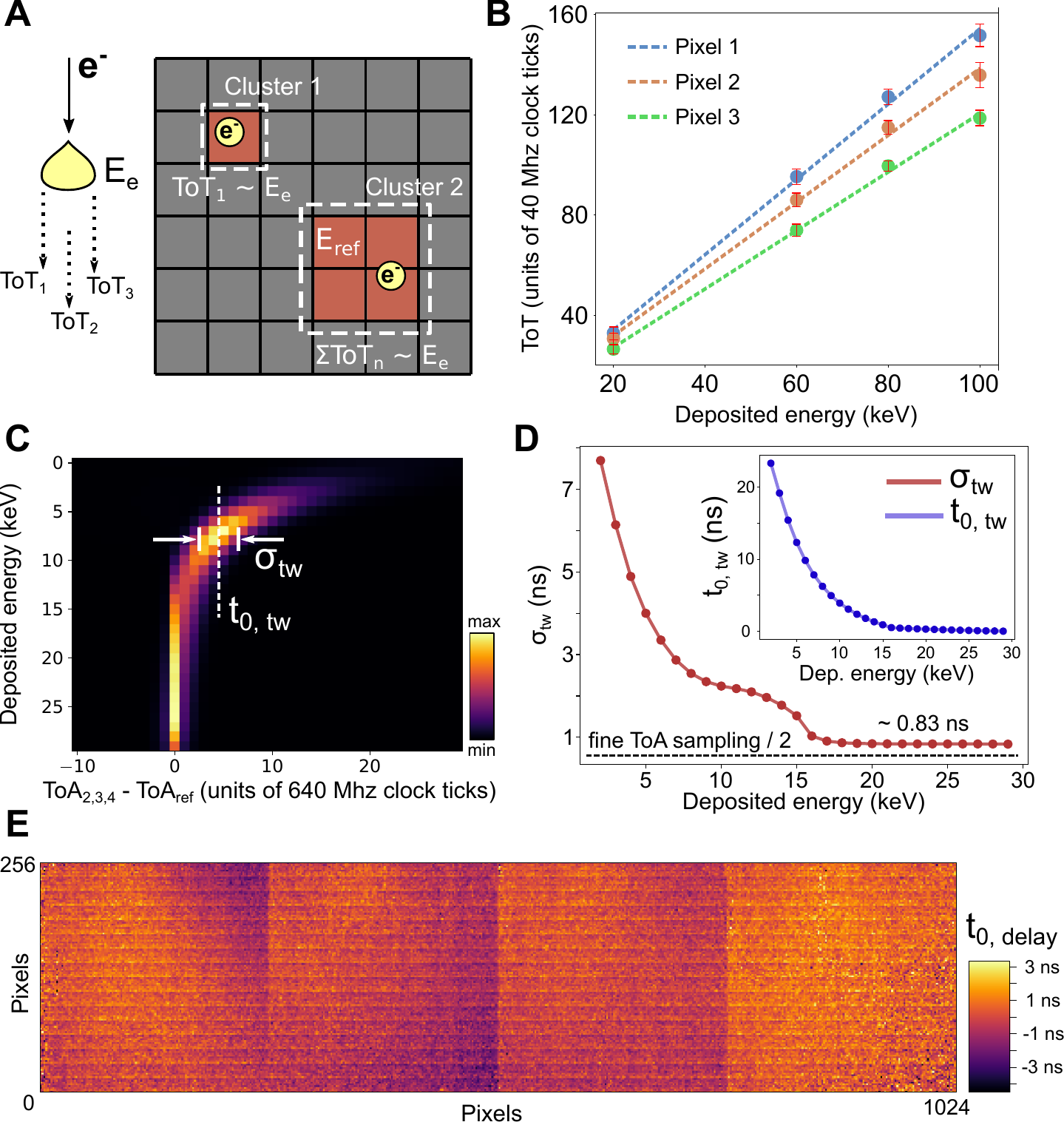}
    \caption{\textbf{Energy, time-walk and delay calibration results}. \textbf{(A)} Energy calibration is done by analyzing clusters with a single hit for multiple incident electron energies ($E_e$) ranging from 20 keV to 100 keV, as shown by cluster 1. For time-walk calibration, exemplified by cluster 2, clusters of 3 and 4 hits containing necessarily an electron hit with energy $E_{ref}$ = 30 keV are used, for an incident electron energy of 60 keV.  \textbf{(B)} The relation between ToT and the hit deposited energy for three distinct pixels. Although there is a linear relationship between the three pixels, the angular and the linear coefficients differ. \textbf{(C)} The time-walk effect integrated along an entire chip array (256 x 256). The hit arrival time at 30 keV is defined as 0, and the relative time is plotted as a function of the deposited energy. For each energy, a Gaussian fit is used to extract the central time ($t_{0, tw}$) and the standard deviation ($\sigma_{tw}$). \textbf{(D)} The values extracted from (C) of the distribution. For deposited energies above 15 keV, the fitted standard deviation is approximately $\sim 0.83$ ns, or roughly half a time bin of 1.5625 ns. \textbf{(E)} The time delay ($t_{0,delay}$) calibration as a function of the detector pixel array, measured by performing temporal coincidences between electrons and photons.}
    \label{Figure2}
\end{figure*}

\subsection*{Energy calibration}
To perform the energy calibration, the electron beam in vacuum is uniformly spread throughout the entire detector for the electron energies of 20 keV, 60 keV, 80 keV, and 100 keV. The cluster identification algorithm is used to sort hits with a unity cluster size, roughly assuring that the energy deposited has not been shared with nearby pixels, and thus allowing the correspondence, pixel-by-pixel, of the deposited energy and the ToT. This is exemplified by the Cluster 1 in Figure \ref{Figure2}A. The histogram of these hits in the pixel array matrix is fitted by Gaussians, and the average ToT per pixel per electron energy is extracted. Figure \ref{Figure2}B shows the result of these means for three different pixels. In the 20 - 100 keV energy range, the relationship of the ToT and electron energy is linear, and a linear fit is used to extract the angular and the offset component per pixel (additional information can be found in the supplementary material, SM). This energy-ToT relationship gives us a glimpse of the data processing of the ASIC, and ultimately allows us to make a better correspondence between the received digitalized ToT and the expected signal amplitude received in the analog input. For even smaller deposited energies, this relationship is no longer linear \cite{pitters2019time} and the deposited energy approaches the pixel threshold value as the ToT approaches zero.

\subsection*{Time-walk \& pixel delay calibration}
To correct the time-walk, a uniformly illuminated detector dataset is used once more. The electron energy is fixed at 60 keV, which provides a good compromise between sufficiently low electron energy to reduce the ionization volume and sufficiently high energy to produce clusters between 1 to 6 hits. To have a controlled dataset, clusters are then post-selected and must have one pixel hit with exactly $E_{ref}=$ 30 keV of deposited energy, and the cluster size must have 3 or 4 hits, as shown by the Cluster 2 in Figure \ref{Figure2}A. The 30 keV hit works as a reference time of arrival value (ToA$_{ref}$) \cite{turecek2016usb}. The other 2 or 3 pixels are used to create a histogram of the electron energy as a function of the time shift between their own time of arrival and the reference value (ToA$_{2, 3, 4} - $ToA$_{ref}$), as illustrated in Figure \ref{Figure2}C for an entire chip array (256 x 256 pixels), promptly exposing the time-walk effect, i.e., the large time differences for low energy electrons. 
The 2D histogram of Figure \ref{Figure2}C is fitted with a Gaussian for every deposited energy, in which the values of the standard deviation ($\sigma_{tw}$) can be seen in Figure \ref{Figure2}D, reaching a constant value of approximately 0.83 ns. The center of the Gaussians ($t_{0, tw}$) is shown in the inset, demonstrating a hyperbolic relationship. To calibrate our data, we have done a similar procedure, but the fitting was also performed pixel-by-pixel, and we have used equation \ref{Equation1} for the interval $5 \leq E < 30$ keV to extract the a, b, and c coefficients, which are later used to correct the raw data. Note that although the time shift can be easily corrected, the standard deviation value represents an intrinsic uncertainty of the instrument in these experimental and data processing conditions. 



There are already many insights on obtaining a relatively good time response with TPX3 with a low-to-none effort on its calibration. Figure \ref{Figure2}C shows the average result for an entire chip array, and, even without a pixel-by-pixel calibration, the deviation of the time-walk reaches $\sim 0.83$ ns, almost half the fine ToA sampling for $E > 15$ keV. 
As mentioned, the effects of time-walk are strongly mitigated by a large charge deposition. Increasing the microscope acceleration voltage is not directly a good option, considering the number of hits per cluster will increase with a big charge sharing between them; additionally, and related, the ionization volume will grow accordingly, increasing the uncertainty of the charge creation. Reducing the TPX3 threshold is, on the contrary, a better reaction. For a given signal amplitude, the time-walk effect will be reduced as the threshold approaches zero, as seen in Figure \ref{Figure1}B. By post-selecting hits with high deposited energies, e.g. higher than 15 keV, the standard deviation is smaller than the ToA sampling and the Gaussian center displacement is less pronounced, as can be seen in the inset of Figure \ref{Figure2}D.

The time-walk calibration discussed above is a relative method, as it only uses ToA values from nearby pixels in the procedure, which leaves unaccounted net propagation delays in the clock signal distribution pixel-by-pixel. To do this calibration, a common reference signal must be used, allowing indirect comparison of this propagation delay. In our case, we have used UV photons and electron correlations by performing CLE experiments in a hexagonal boron nitride sample (\textit{h}--BN). The photons are sent to a single photon-counting PMT and coincident histograms are plotted as a function of the pixel matrix coordinate, and the center position of the Gaussian ($t_{0, delay}$) fit provides the propagation delay values. The obtained delay calibration array is shown in Figure \ref{Figure2}E. Further details are also present in the SM of this work.

\subsection*{Impact of the calibration using electron-photon temporal correlations}

\begin{figure*}[t]
    \centering
    \includegraphics[width=0.6\textwidth]{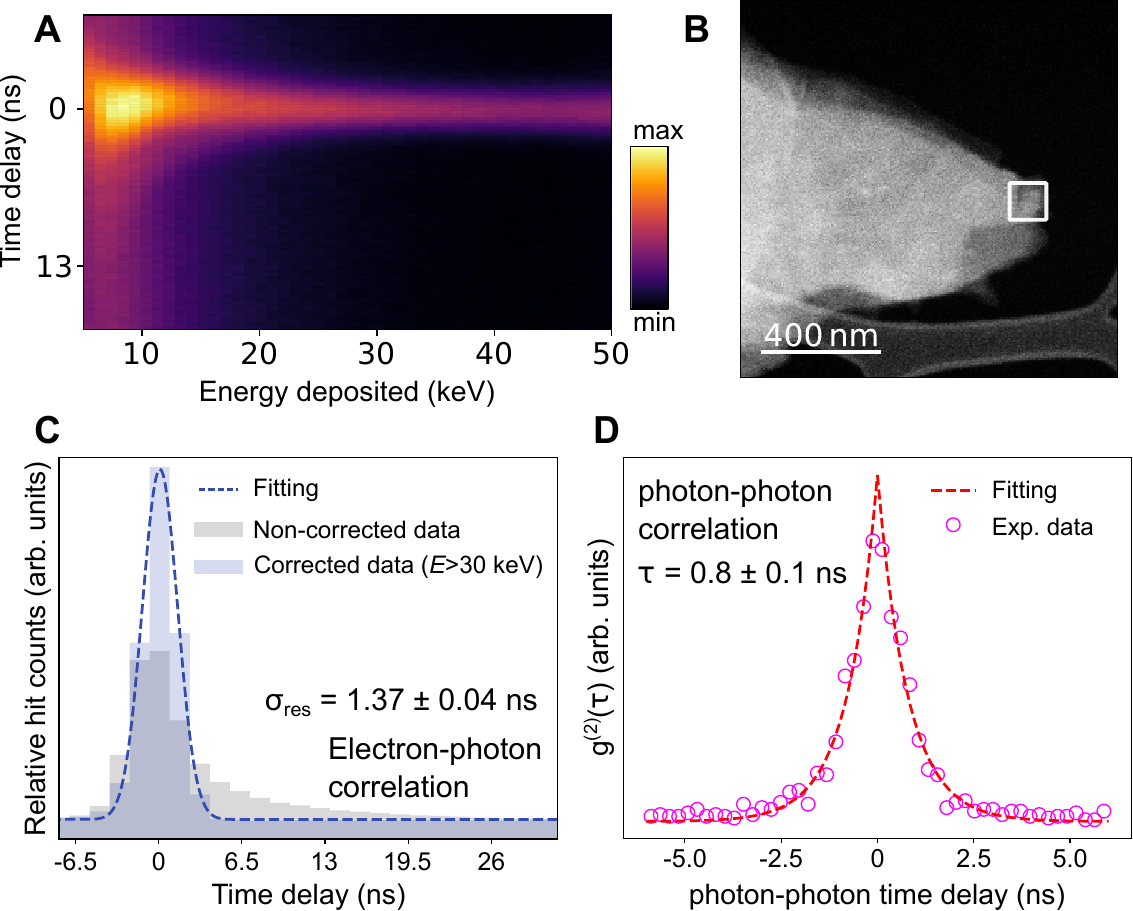}
    \caption{\textbf{The delay calibration and the impact of the described correction methods accessed by electron-photon coincidences}. (\textbf{A}) The 2D histogram of the time delay between electron-photon pairs as a function of the energy deposited in the pixel hit after the time-walk and the pixel delay calibration. (\textbf{B}) The high-angular dark-field image of the used \textit{h}--BN flake. Data has been averaged by scanning in the white rectangle area of approximately 125 x 125 nm$^2$. (\textbf{C}) The time distributions for the non-corrected data, with a long tail towards the positive time delay direction. After the calibration and post-selecting hits with $> 30$ keV deposited energy, the curve approaches a Gaussian distribution with a fitted sigma of $1.37 \pm 0.04$ ns. (\textbf{D}) The photon-photon correlation curve done by HBT interferometry. The curve is plotted by a exponential decay symmetric with respect to the zero time delay. The obtained value from the fitting is $\tau = 0.8 \pm 0.1$ ns.}
    \label{Figure3}
\end{figure*}

Figure \ref{Figure3}A shows experimental results of the time delay between a photon and an electron as a function of the deposited energy after the time-walk and the delay calibration averaged through all the pixels. Measurements were taken in a \textit{h}--BN flake in a region of approximately 125 x 125 nm$^2$, highlighted by the white rectangle in the annular dark-field image of the sample, as shown in Figure \ref{Figure3}B. As we are interested in the averaged temporal resolution throughout the entire pixel matrix, the electron beam has been rastered in the Timepix3 detector in order to increase the pixel occupancy. Post-selecting high energetic hits ($E \geq 30$ keV) after the time-walk and delay calibration produce the best possible detector's time response, showing a standard deviation of $\sigma_{res} = 1.37 \pm 0.04$ ns, smaller than the bin width of the electron ToA fine timestamping. From Figure \ref{Figure3}A, we can see that there are more hits with low deposited energy ($E < 15$ keV). However, they are usually associated with a high-energy hit within the same cluster, meaning that data loss after hit post-selection is not too critical. Finally, it is important to note that \textit{h}-BN sample's lifetime is convoluted in this result. To discern this contribution, we have performed HBT interferometry at the sample, temporally correlating two photons instead of one electron and one photon. The results are shown in Figure \ref{Figure3}D, and the decay's lifetime has been determined as $\tau = 0.8 \pm 0.1$ ns. As recently demonstrated \cite{varkentina2023excitations}, the \textit{h}--BN lifetime can be seen in our electron-photon correlations by fitting exponential decays in both sides of the time delay curves, and further discussions can be found in the SM.



\subsection*{Cosmic rays tracks}

\begin{figure*}[t!]
    \centering
    \includegraphics[width=0.7\textwidth]{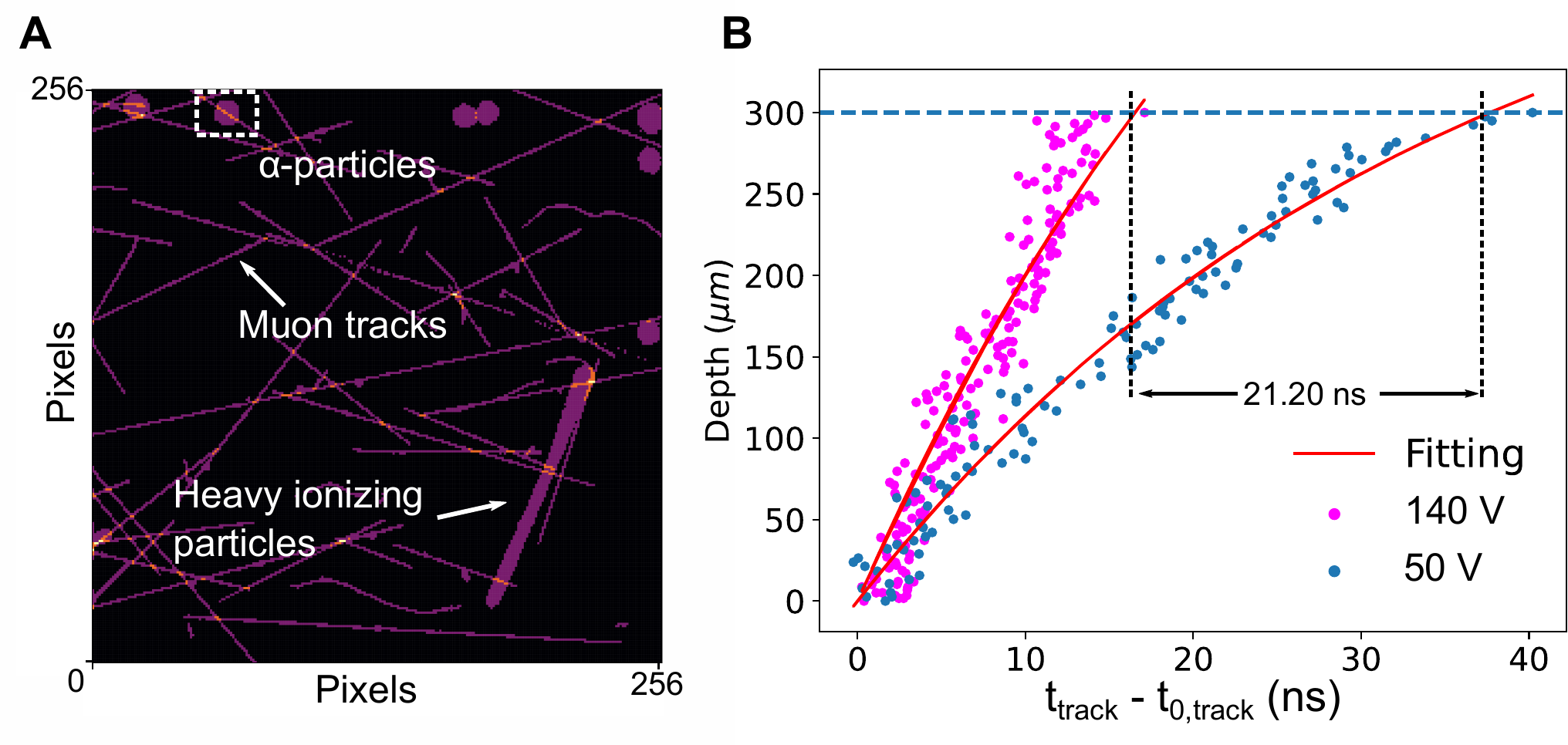}
    \caption{\textbf{Cosmic rays tracks can provide further insights in the charge dynamics of Timepix3}. \textbf{(A)} Example of different ionization particles that can hit the TPX3 detector. Data is filtered by post-selecting clusters with more than 80 hits. Muon tracks are thin and can cross the 300 $\mu m$ sensor thickness. The start and end point's pixel values allow us to determine its trajectory inside the sensor layer. \textbf{(B)} Muon tracks analyzed for two different detector biases. A larger bias, and hence a larger electric field, permit the charges to be collected faster. The fitting uses a simple charge drift model \cite{bergmann20173d}.}
    \label{Figure4}
\end{figure*}

Multiple ionizing particles can hit the TPX3 detector during data acquisition. These produce a variety of shapes and sizes, as can be see in Figure \ref{Figure4}A. Large blobs are typically associated with heavy and short-range ionizing particles, such as $\alpha$ particles. When these heavy tracks are elongated, they are typically associated with protons or atom nuclei \cite{heijne2013measuring, granja2018directional}. More interesting for this work are highly energetic ($\sim$ GeV) and light particles such as muons. As these particles entirely cross the sensor layer, it is possible to identify their precise path by the initial and final pixel position values and by the detector thickness (300 $\mu $m in our case), as illustrated in Figure \ref{Figure1}A. The charge collection dynamics can thus be studied by the obtained values of the ToA $t_{track}$, in which the first detected charge is taken as a reference value $t_{0, track}$. Two of these tracks with cluster sizes greater than 150 hits are shown in Figure \ref{Figure4}B for a detector's bias voltage of 140 V and 50 V. By changing from 140 V to 50 V bias, charges created at the surface of the sensor layer arrive 21.20 ns later, determined by fitting a charge drift model \cite{bergmann20173d} to the experimental data.

\begin{figure}[t!]
    \centering
    \includegraphics[width=0.35\textwidth]{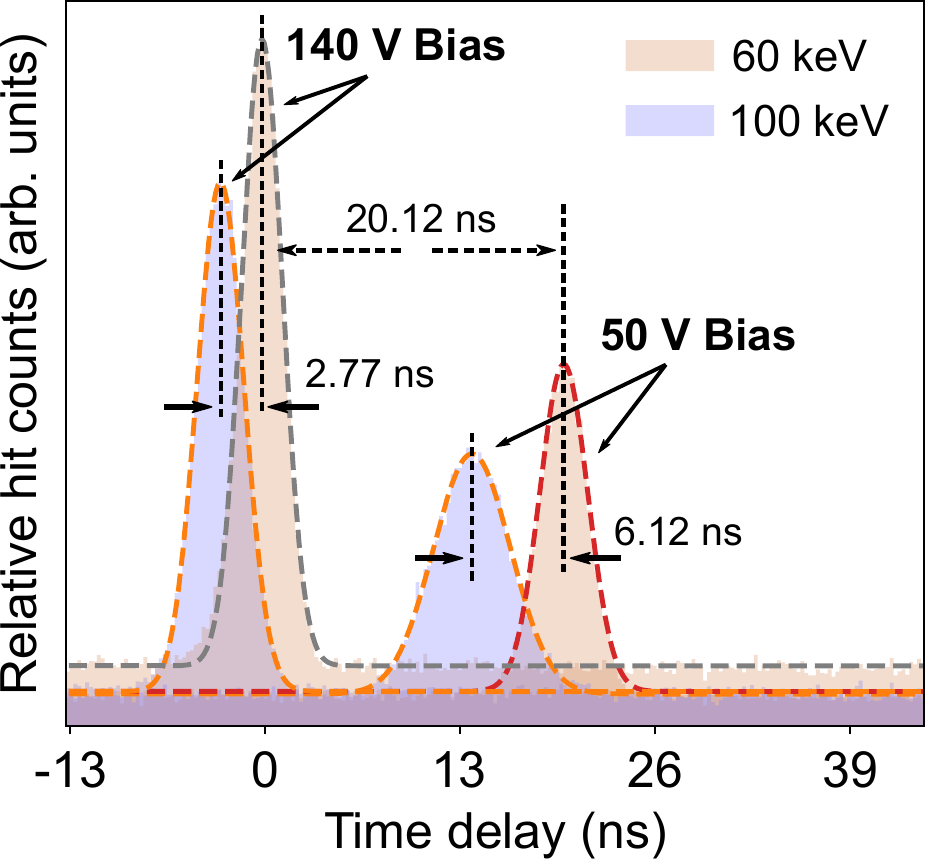}
    \caption{\textbf{Electron-photon temporal correlation as a function of the electron energy and the Timepix3 voltage bias}. Performing photon-electron coincidences for 60 keV and 100 keV acceleration voltages for 50 V and 140 V detector's bias. Electrons with 100 keV reach the ASIC earlier but have a worsened temporal resolution. Reducing the detector's bias delays the charge arrival time and degrades the detector's temporal resolution.}
    \label{Figure5}
\end{figure}

Additionally, this model can be confronted with photon-electron correlation measurements. Between the two measured biases, the time delay between the photon and electron correlation increases by roughly 20.12 ns for the 50 V bias, as shown in Figure \ref{Figure5}. The value is slightly smaller than the 21.20 ns measured from the muon track, and the observed difference is presumably due to the fact that charges created from fast electrons are not precisely induced at the surface of the sensor layer but rather at a few microns inside the bulk Si material. To sustain this, we have performed experiments between 60 keV and 100 keV acceleration voltages. For both biases, the slower electron arrives later, which is expected considering the reasoning that slower electrons are absorbed closer to the sensor surface. Equally interesting, the observed standard deviation for the Gaussian fitting of the curves in Figure \ref{Figure5} changes significantly. Due to the reduced ionization volume, smaller electron voltages produce smaller standard deviations. Analogously, a reduced bias also degrades the temporal resolution by the more significant skewness of the charge collection curve (Figure \ref{Figure4}B). 
For 60 keV at 140 V bias, the standard deviation is $1.37 \pm 0.04$ ns. If the acceleration voltage is 100 keV, this value increases to $1.56 \pm 0.04$ ns. For the 50 V bias, these values are $2.57 \pm 0.06$ ns and $1.61 \pm 0.04$ ns for 100 keV and 60 keV, respectively.

Finally, the tools above allow us to try to estimate the achievable temporal resolution as a function of the electron energy, the sensor thickness, and the applied voltage, provided that the temporal calibration is correctly performed. For this, we have used a Monte Carlo simulation software, CASINO \cite{hovington1997casino}, to study the spatial distribution of the deposited energy when fast electrons hit a silicon slab. The reference values were extracted from the silicon slab depth in which the cathodoluminescence probability is maximum. These values are roughly 11 $\mu$m, 24 $\mu$m, 74 $\mu$m, and 140 $\mu$m for 60 keV, 100 keV, 200 keV, and 300 keV respectively, and the corresponding uncertainties are 0.8 ns, 1.7 ns, 5.0 ns, and 8.8 ns, all of them considering a 140 V detector's bias voltage. Indeed, this estimate is very simplistic, and further analysis must be performed to retrieve more accurate values. For this, better well-suited Monte Carlo toolkits must be used, such as Geant4 \cite{agostinelli2003geant4}, actively developed by CERN for particle-matter interaction simulation and detector development, in which more recent frameworks consist of a complete simulation of hybrid-pixel detectors, including the charge transport dynamics, the pre-amplifier response, and the expected values of ToA and ToT \cite{schubel2014geant4}.

\section*{Conclusions \& perspectives}

In this work, we have applied well-known but also developed new tools for time calibration of the Timepix3 HPD in the context of electron microscopy. In particular, we have accounted for the energy calibration, the time-walk effect, and the time delay between the pixel array matrix. Additionally, we have shown how photon-electron coincidence events can help the calibration but also to verify the impact of previous steps in the final processed data. Further, we have used highly energetic cosmic rays tracks to unveil the charge deposition mechanism experimentally under different sensor biasing voltages. The obtained values were confronted with the photon-electron coincidence experiments, showing a remarkable similarity between the obtained values. With these experiments, we were able to show that higher energetic electrons produce charges deeply in the sensor layer because of the reduced drift time but also degrade the maximum attainable temporal resolution probably due to the increased ionization volume. Unfortunately, the microscope used has a maximum acceleration voltage of 100 keV, undermining further investigation, and more systematic studies must be performed to confirm if indeed, the ionization volume can not be corrected. Finally, we have deduced from the experiments as mentioned above that the uncertainty related to the ionization volume is approximately 0.8 ns at 60 keV electrons, while this value increases to 1.7 ns for 100 keV electrons, to 5.0 ns for 200 keV, and 8.8 ns to 300 keV electrons.

Although we have presented a relatively easy calibration method, this is far from ideal. Because our time-walk calibration depends on a uniform electron illumination, the obtained time intervals have contributions from both the charge dynamics in the sensor layer and the digital time conversion provided by the ASIC. A better way of calibration is to rely on test pulsing \cite{poikela2014timepix3, pitters2019time}, which depends solely on the ASIC, and then afterward perform what has been described in our work to account and access residual charge dynamics contributions. Additionally, the time delay calibration procedure here depends on a very large data acquisition and is prone to uncertainties if the lifetime of the material is comparable to the expected maximum attainable temporal resolution, which in our work have been measured by HBT interferometry \cite{meuret2015photon}. An interesting solution for the time delay calibration is using ultrafast electron microscopes, in which electron pulses with sub-picosecond temporal resolution are routinely achieved \cite{feist2017ultrafast, houdellier2018development}. A final but significant source of uncertainty is due to our energy calibration measurement, in which the low energy region ($< 20$ keV) has been considered linear although this is not correct.


Timepix4, the successor of Timepix3 capable of achieving sub 200 ps time binning is already under tests \cite{campbell2016towards, llopart2022timepix4, heijhoff2022timing}. It would be able to be operated in event-based or frame-based mode. In this later, the 16-bit counter will provide the necessary electron dynamics to be able to potentially establish itself as a standard electron detector in many electron microscopes. Although the expected $\sim$ 100 ps temporal resolution may not be directly feasible in electron microscopy for reasons already discussed in this work, the much higher expected data flux will be able to perform a plethora of experiments without worrying about electron beam saturation \cite{llopart2022timepix4}. The theoretical total readout bandwidth of such a detector can reach as high as 160 Gbps of data transfer, which will definitely trigger not only a new way of data storage but also different ways of interfacing them with the electron microscope.


\section*{Acknowledgements}
The present project has received funding from the European
Union’s Horizon 2020 research and nnovation programme un-
dergrant agreement No 823717 (ESTEEM3) and 101017720
(EBEAM). Amsterdam Scientific Instruments (ASI) is acknowledged for many fruitful technical discussions.
\bibliography{newbiblio.bib}

\begin{thebibliography}{32}%
\makeatletter
\providecommand \@ifxundefined [1]{%
 \@ifx{#1\undefined}
}%
\providecommand \@ifnum [1]{%
 \ifnum #1\expandafter \@firstoftwo
 \else \expandafter \@secondoftwo
 \fi
}%
\providecommand \@ifx [1]{%
 \ifx #1\expandafter \@firstoftwo
 \else \expandafter \@secondoftwo
 \fi
}%
\providecommand \natexlab [1]{#1}%
\providecommand \enquote  [1]{``#1''}%
\providecommand \bibnamefont  [1]{#1}%
\providecommand \bibfnamefont [1]{#1}%
\providecommand \citenamefont [1]{#1}%
\providecommand \href@noop [0]{\@secondoftwo}%
\providecommand \href [0]{\begingroup \@sanitize@url \@href}%
\providecommand \@href[1]{\@@startlink{#1}\@@href}%
\providecommand \@@href[1]{\endgroup#1\@@endlink}%
\providecommand \@sanitize@url [0]{\catcode `\\12\catcode `\$12\catcode
  `\&12\catcode `\#12\catcode `\^12\catcode `\_12\catcode `\%12\relax}%
\providecommand \@@startlink[1]{}%
\providecommand \@@endlink[0]{}%
\providecommand \url  [0]{\begingroup\@sanitize@url \@url }%
\providecommand \@url [1]{\endgroup\@href {#1}{\urlprefix }}%
\providecommand \urlprefix  [0]{URL }%
\providecommand \Eprint [0]{\href }%
\providecommand \doibase [0]{https://doi.org/}%
\providecommand \selectlanguage [0]{\@gobble}%
\providecommand \bibinfo  [0]{\@secondoftwo}%
\providecommand \bibfield  [0]{\@secondoftwo}%
\providecommand \translation [1]{[#1]}%
\providecommand \BibitemOpen [0]{}%
\providecommand \bibitemStop [0]{}%
\providecommand \bibitemNoStop [0]{.\EOS\space}%
\providecommand \EOS [0]{\spacefactor3000\relax}%
\providecommand \BibitemShut  [1]{\csname bibitem#1\endcsname}%
\let\auto@bib@innerbib\@empty
\bibitem [{\citenamefont {van Schayck}\ \emph {et~al.}(2020)\citenamefont {van
  Schayck}, \citenamefont {van Genderen}, \citenamefont {Maddox}, \citenamefont
  {Roussel}, \citenamefont {Boulanger}, \citenamefont {Fr{\"o}jdh},
  \citenamefont {Abrahams}, \citenamefont {Peters},\ and\ \citenamefont
  {Ravelli}}]{van2020sub}%
  \BibitemOpen
  \bibfield  {author} {\bibinfo {author} {\bibfnamefont {J.~P.}\ \bibnamefont
  {van Schayck}}, \bibinfo {author} {\bibfnamefont {E.}~\bibnamefont {van
  Genderen}}, \bibinfo {author} {\bibfnamefont {E.}~\bibnamefont {Maddox}},
  \bibinfo {author} {\bibfnamefont {L.}~\bibnamefont {Roussel}}, \bibinfo
  {author} {\bibfnamefont {H.}~\bibnamefont {Boulanger}}, \bibinfo {author}
  {\bibfnamefont {E.}~\bibnamefont {Fr{\"o}jdh}}, \bibinfo {author}
  {\bibfnamefont {J.-P.}\ \bibnamefont {Abrahams}}, \bibinfo {author}
  {\bibfnamefont {P.~J.}\ \bibnamefont {Peters}},\ and\ \bibinfo {author}
  {\bibfnamefont {R.~B.}\ \bibnamefont {Ravelli}},\ }\bibfield  {title}
  {\bibinfo {title} {Sub-pixel electron detection using a convolutional neural
  network},\ }\href {https://doi.org/10.1016/j.ultramic.2020.113091} {\bibfield
   {journal} {\bibinfo  {journal} {Ultramicroscopy}\ }\textbf {\bibinfo
  {volume} {218}},\ \bibinfo {pages} {113091} (\bibinfo {year}
  {2020})}\BibitemShut {NoStop}%
\bibitem [{\citenamefont {Jannis}\ \emph {et~al.}(2022)\citenamefont {Jannis},
  \citenamefont {Hofer}, \citenamefont {Gao}, \citenamefont {Xie},
  \citenamefont {B{\'e}ch{\'e}}, \citenamefont {Pennycook},\ and\ \citenamefont
  {Verbeeck}}]{jannis2022event}%
  \BibitemOpen
  \bibfield  {author} {\bibinfo {author} {\bibfnamefont {D.}~\bibnamefont
  {Jannis}}, \bibinfo {author} {\bibfnamefont {C.}~\bibnamefont {Hofer}},
  \bibinfo {author} {\bibfnamefont {C.}~\bibnamefont {Gao}}, \bibinfo {author}
  {\bibfnamefont {X.}~\bibnamefont {Xie}}, \bibinfo {author} {\bibfnamefont
  {A.}~\bibnamefont {B{\'e}ch{\'e}}}, \bibinfo {author} {\bibfnamefont {T.~J.}\
  \bibnamefont {Pennycook}},\ and\ \bibinfo {author} {\bibfnamefont
  {J.}~\bibnamefont {Verbeeck}},\ }\bibfield  {title} {\bibinfo {title} {Event
  driven {4D} {STEM} acquisition with a {T}imepix3 detector: microsecond dwell
  time and faster scans for high precision and low dose applications},\ }\href
  {https://doi.org/10.1016/j.ultramic.2021.113423} {\bibfield  {journal}
  {\bibinfo  {journal} {Ultramicroscopy}\ }\textbf {\bibinfo {volume} {233}},\
  \bibinfo {pages} {113423} (\bibinfo {year} {2022})}\BibitemShut {NoStop}%
\bibitem [{\citenamefont {Hart}\ \emph {et~al.}(2017)\citenamefont {Hart},
  \citenamefont {Lang}, \citenamefont {Leff}, \citenamefont {Longo},
  \citenamefont {Trevor}, \citenamefont {Twesten},\ and\ \citenamefont
  {Taheri}}]{hart2017direct}%
  \BibitemOpen
  \bibfield  {author} {\bibinfo {author} {\bibfnamefont {J.~L.}\ \bibnamefont
  {Hart}}, \bibinfo {author} {\bibfnamefont {A.~C.}\ \bibnamefont {Lang}},
  \bibinfo {author} {\bibfnamefont {A.~C.}\ \bibnamefont {Leff}}, \bibinfo
  {author} {\bibfnamefont {P.}~\bibnamefont {Longo}}, \bibinfo {author}
  {\bibfnamefont {C.}~\bibnamefont {Trevor}}, \bibinfo {author} {\bibfnamefont
  {R.~D.}\ \bibnamefont {Twesten}},\ and\ \bibinfo {author} {\bibfnamefont
  {M.~L.}\ \bibnamefont {Taheri}},\ }\bibfield  {title} {\bibinfo {title}
  {Direct detection electron energy-loss spectroscopy: a method to push the
  limits of resolution and sensitivity},\ }\href
  {https://doi.org/10.1038/s41598-017-07709-4} {\bibfield  {journal} {\bibinfo
  {journal} {Scientific reports}\ }\textbf {\bibinfo {volume} {7}},\ \bibinfo
  {pages} {1} (\bibinfo {year} {2017})}\BibitemShut {NoStop}%
\bibitem [{\citenamefont {Tenc{\'e}}\ \emph {et~al.}(2020)\citenamefont
  {Tenc{\'e}}, \citenamefont {Blazit}, \citenamefont {Li}, \citenamefont
  {Krajnak}, \citenamefont {del Busto}, \citenamefont {Skogeby}, \citenamefont
  {Cambou}, \citenamefont {Kociak}, \citenamefont {Stephan},\ and\
  \citenamefont {Gloter}}]{tence2020electron}%
  \BibitemOpen
  \bibfield  {author} {\bibinfo {author} {\bibfnamefont {M.}~\bibnamefont
  {Tenc{\'e}}}, \bibinfo {author} {\bibfnamefont {J.-D.}\ \bibnamefont
  {Blazit}}, \bibinfo {author} {\bibfnamefont {X.}~\bibnamefont {Li}}, \bibinfo
  {author} {\bibfnamefont {M.}~\bibnamefont {Krajnak}}, \bibinfo {author}
  {\bibfnamefont {E.~N.}\ \bibnamefont {del Busto}}, \bibinfo {author}
  {\bibfnamefont {R.}~\bibnamefont {Skogeby}}, \bibinfo {author} {\bibfnamefont
  {L.}~\bibnamefont {Cambou}}, \bibinfo {author} {\bibfnamefont
  {M.}~\bibnamefont {Kociak}}, \bibinfo {author} {\bibfnamefont
  {O.}~\bibnamefont {Stephan}},\ and\ \bibinfo {author} {\bibfnamefont
  {A.}~\bibnamefont {Gloter}},\ }\bibfield  {title} {\bibinfo {title} {Electron
  energy-loss spectroscopy using merlinem-medipix3 detector},\ }\href
  {https://doi.org/https://doi.org/10.1017/S1431927620019881} {\bibfield
  {journal} {\bibinfo  {journal} {Microscopy and Microanalysis}\ }\textbf
  {\bibinfo {volume} {26}},\ \bibinfo {pages} {1940} (\bibinfo {year}
  {2020})}\BibitemShut {NoStop}%
\bibitem [{\citenamefont {Auad}\ \emph
  {et~al.}(2022{\natexlab{a}})\citenamefont {Auad}, \citenamefont {Walls},
  \citenamefont {Blazit}, \citenamefont {St{\'e}phan}, \citenamefont {Tizei},
  \citenamefont {Kociak}, \citenamefont {de~La~Pe{\~n}a},\ and\ \citenamefont
  {Tenc{\'e}}}]{auad2022event}%
  \BibitemOpen
  \bibfield  {author} {\bibinfo {author} {\bibfnamefont {Y.}~\bibnamefont
  {Auad}}, \bibinfo {author} {\bibfnamefont {M.}~\bibnamefont {Walls}},
  \bibinfo {author} {\bibfnamefont {J.-D.}\ \bibnamefont {Blazit}}, \bibinfo
  {author} {\bibfnamefont {O.}~\bibnamefont {St{\'e}phan}}, \bibinfo {author}
  {\bibfnamefont {L.~H.}\ \bibnamefont {Tizei}}, \bibinfo {author}
  {\bibfnamefont {M.}~\bibnamefont {Kociak}}, \bibinfo {author} {\bibfnamefont
  {F.}~\bibnamefont {de~La~Pe{\~n}a}},\ and\ \bibinfo {author} {\bibfnamefont
  {M.}~\bibnamefont {Tenc{\'e}}},\ }\bibfield  {title} {\bibinfo {title}
  {Event-based hyperspectral eels: towards nanosecond temporal resolution},\
  }\href {https://doi.org/10.1016/j.ultramic.2022.113539} {\bibfield  {journal}
  {\bibinfo  {journal} {Ultramicroscopy}\ }\textbf {\bibinfo {volume} {239}},\
  \bibinfo {pages} {113539} (\bibinfo {year} {2022}{\natexlab{a}})}\BibitemShut
  {NoStop}%
\bibitem [{\citenamefont {Ballabriga}\ \emph {et~al.}(2018)\citenamefont
  {Ballabriga}, \citenamefont {Campbell},\ and\ \citenamefont
  {Llopart}}]{ballabriga2018asic}%
  \BibitemOpen
  \bibfield  {author} {\bibinfo {author} {\bibfnamefont {R.}~\bibnamefont
  {Ballabriga}}, \bibinfo {author} {\bibfnamefont {M.}~\bibnamefont
  {Campbell}},\ and\ \bibinfo {author} {\bibfnamefont {X.}~\bibnamefont
  {Llopart}},\ }\bibfield  {title} {\bibinfo {title} {Asic developments for
  radiation imaging applications: The {M}edipix and {T}imepix family},\ }\href
  {https://doi.org/10.1016/j.nima.2017.07.029} {\bibfield  {journal} {\bibinfo
  {journal} {Nuclear Instruments and Methods in Physics Research Section A:
  Accelerators, Spectrometers, Detectors and Associated Equipment}\ }\textbf
  {\bibinfo {volume} {878}},\ \bibinfo {pages} {10} (\bibinfo {year}
  {2018})}\BibitemShut {NoStop}%
\bibitem [{\citenamefont {Poikela}\ \emph {et~al.}(2014)\citenamefont
  {Poikela}, \citenamefont {Plosila}, \citenamefont {Westerlund}, \citenamefont
  {Campbell}, \citenamefont {De~Gaspari}, \citenamefont {Llopart},
  \citenamefont {Gromov}, \citenamefont {Kluit}, \citenamefont {Van~Beuzekom},
  \citenamefont {Zappon} \emph {et~al.}}]{poikela2014timepix3}%
  \BibitemOpen
  \bibfield  {author} {\bibinfo {author} {\bibfnamefont {T.}~\bibnamefont
  {Poikela}}, \bibinfo {author} {\bibfnamefont {J.}~\bibnamefont {Plosila}},
  \bibinfo {author} {\bibfnamefont {T.}~\bibnamefont {Westerlund}}, \bibinfo
  {author} {\bibfnamefont {M.}~\bibnamefont {Campbell}}, \bibinfo {author}
  {\bibfnamefont {M.}~\bibnamefont {De~Gaspari}}, \bibinfo {author}
  {\bibfnamefont {X.}~\bibnamefont {Llopart}}, \bibinfo {author} {\bibfnamefont
  {V.}~\bibnamefont {Gromov}}, \bibinfo {author} {\bibfnamefont
  {R.}~\bibnamefont {Kluit}}, \bibinfo {author} {\bibfnamefont
  {M.}~\bibnamefont {Van~Beuzekom}}, \bibinfo {author} {\bibfnamefont
  {F.}~\bibnamefont {Zappon}}, \emph {et~al.},\ }\bibfield  {title} {\bibinfo
  {title} {{T}imepix3: a 65k channel hybrid pixel readout chip with
  simultaneous toa/tot and sparse readout},\ }\href
  {https://doi.org/10.1088/1748-0221/9/05/C05013} {\bibfield  {journal}
  {\bibinfo  {journal} {Journal of instrumentation}\ }\textbf {\bibinfo
  {volume} {9}},\ \bibinfo {pages} {C05013} (\bibinfo {year}
  {2014})}\BibitemShut {NoStop}%
\bibitem [{\citenamefont {Stevens}\ \emph {et~al.}(2018)\citenamefont
  {Stevens}, \citenamefont {Luzi}, \citenamefont {Yang}, \citenamefont
  {Kovarik}, \citenamefont {Mehdi}, \citenamefont {Liyu}, \citenamefont
  {Gehm},\ and\ \citenamefont {Browning}}]{stevens2018sub}%
  \BibitemOpen
  \bibfield  {author} {\bibinfo {author} {\bibfnamefont {A.}~\bibnamefont
  {Stevens}}, \bibinfo {author} {\bibfnamefont {L.}~\bibnamefont {Luzi}},
  \bibinfo {author} {\bibfnamefont {H.}~\bibnamefont {Yang}}, \bibinfo {author}
  {\bibfnamefont {L.}~\bibnamefont {Kovarik}}, \bibinfo {author} {\bibfnamefont
  {B.}~\bibnamefont {Mehdi}}, \bibinfo {author} {\bibfnamefont
  {A.}~\bibnamefont {Liyu}}, \bibinfo {author} {\bibfnamefont {M.}~\bibnamefont
  {Gehm}},\ and\ \bibinfo {author} {\bibfnamefont {N.}~\bibnamefont
  {Browning}},\ }\bibfield  {title} {\bibinfo {title} {A sub-sampled approach
  to extremely low-dose {STEM}},\ }\href {https://doi.org/10.1063/1.5016192}
  {\bibfield  {journal} {\bibinfo  {journal} {Applied Physics Letters}\
  }\textbf {\bibinfo {volume} {112}},\ \bibinfo {pages} {043104} (\bibinfo
  {year} {2018})}\BibitemShut {NoStop}%
\bibitem [{\citenamefont {Zobelli}\ \emph {et~al.}(2020)\citenamefont
  {Zobelli}, \citenamefont {Woo}, \citenamefont {Tararan}, \citenamefont
  {Tizei}, \citenamefont {Brun}, \citenamefont {Li}, \citenamefont
  {St{\'e}phan}, \citenamefont {Kociak},\ and\ \citenamefont
  {Tenc{\'e}}}]{zobelli2020spatial}%
  \BibitemOpen
  \bibfield  {author} {\bibinfo {author} {\bibfnamefont {A.}~\bibnamefont
  {Zobelli}}, \bibinfo {author} {\bibfnamefont {S.~Y.}\ \bibnamefont {Woo}},
  \bibinfo {author} {\bibfnamefont {A.}~\bibnamefont {Tararan}}, \bibinfo
  {author} {\bibfnamefont {L.~H.}\ \bibnamefont {Tizei}}, \bibinfo {author}
  {\bibfnamefont {N.}~\bibnamefont {Brun}}, \bibinfo {author} {\bibfnamefont
  {X.}~\bibnamefont {Li}}, \bibinfo {author} {\bibfnamefont {O.}~\bibnamefont
  {St{\'e}phan}}, \bibinfo {author} {\bibfnamefont {M.}~\bibnamefont
  {Kociak}},\ and\ \bibinfo {author} {\bibfnamefont {M.}~\bibnamefont
  {Tenc{\'e}}},\ }\bibfield  {title} {\bibinfo {title} {Spatial and spectral
  dynamics in {STEM} hyperspectral imaging using random scan patterns},\ }\href
  {https://doi.org/10.1016/j.ultramic.2019.112912} {\bibfield  {journal}
  {\bibinfo  {journal} {Ultramicroscopy}\ }\textbf {\bibinfo {volume} {212}},\
  \bibinfo {pages} {112912} (\bibinfo {year} {2020})}\BibitemShut {NoStop}%
\bibitem [{\citenamefont {Varkentina}\ \emph {et~al.}(2022)\citenamefont
  {Varkentina}, \citenamefont {Auad}, \citenamefont {Woo}, \citenamefont
  {Zobelli}, \citenamefont {Bocher}, \citenamefont {Blazit}, \citenamefont
  {Li}, \citenamefont {Tenc{\'e}}, \citenamefont {Watanabe}, \citenamefont
  {Taniguchi} \emph {et~al.}}]{varkentina2022cathodoluminescence}%
  \BibitemOpen
  \bibfield  {author} {\bibinfo {author} {\bibfnamefont {N.}~\bibnamefont
  {Varkentina}}, \bibinfo {author} {\bibfnamefont {Y.}~\bibnamefont {Auad}},
  \bibinfo {author} {\bibfnamefont {S.~Y.}\ \bibnamefont {Woo}}, \bibinfo
  {author} {\bibfnamefont {A.}~\bibnamefont {Zobelli}}, \bibinfo {author}
  {\bibfnamefont {L.}~\bibnamefont {Bocher}}, \bibinfo {author} {\bibfnamefont
  {J.-D.}\ \bibnamefont {Blazit}}, \bibinfo {author} {\bibfnamefont
  {X.}~\bibnamefont {Li}}, \bibinfo {author} {\bibfnamefont {M.}~\bibnamefont
  {Tenc{\'e}}}, \bibinfo {author} {\bibfnamefont {K.}~\bibnamefont {Watanabe}},
  \bibinfo {author} {\bibfnamefont {T.}~\bibnamefont {Taniguchi}}, \emph
  {et~al.},\ }\bibfield  {title} {\bibinfo {title} {Cathodoluminescence
  excitation spectroscopy: Nanoscale imaging of excitation pathways},\ }\href
  {https://doi.org/10.1126/sciadv.abq4947} {\bibfield  {journal} {\bibinfo
  {journal} {Science Advances}\ }\textbf {\bibinfo {volume} {8}},\ \bibinfo
  {pages} {eabq4947} (\bibinfo {year} {2022})}\BibitemShut {NoStop}%
\bibitem [{\citenamefont {Feist}\ \emph {et~al.}(2022)\citenamefont {Feist},
  \citenamefont {Huang}, \citenamefont {Arend}, \citenamefont {Yang},
  \citenamefont {Henke}, \citenamefont {Raja}, \citenamefont {Kappert},
  \citenamefont {Wang}, \citenamefont {Louren{\c{c}}o-Martins}, \citenamefont
  {Qiu} \emph {et~al.}}]{feist2022cavity}%
  \BibitemOpen
  \bibfield  {author} {\bibinfo {author} {\bibfnamefont {A.}~\bibnamefont
  {Feist}}, \bibinfo {author} {\bibfnamefont {G.}~\bibnamefont {Huang}},
  \bibinfo {author} {\bibfnamefont {G.}~\bibnamefont {Arend}}, \bibinfo
  {author} {\bibfnamefont {Y.}~\bibnamefont {Yang}}, \bibinfo {author}
  {\bibfnamefont {J.-W.}\ \bibnamefont {Henke}}, \bibinfo {author}
  {\bibfnamefont {A.~S.}\ \bibnamefont {Raja}}, \bibinfo {author}
  {\bibfnamefont {F.~J.}\ \bibnamefont {Kappert}}, \bibinfo {author}
  {\bibfnamefont {R.~N.}\ \bibnamefont {Wang}}, \bibinfo {author}
  {\bibfnamefont {H.}~\bibnamefont {Louren{\c{c}}o-Martins}}, \bibinfo {author}
  {\bibfnamefont {Z.}~\bibnamefont {Qiu}}, \emph {et~al.},\ }\bibfield  {title}
  {\bibinfo {title} {Cavity-mediated electron-photon pairs},\ }\href
  {https://doi.org/10.1126/science.abo5037} {\bibfield  {journal} {\bibinfo
  {journal} {Science}\ }\textbf {\bibinfo {volume} {377}},\ \bibinfo {pages}
  {777} (\bibinfo {year} {2022})}\BibitemShut {NoStop}%
\bibitem [{\citenamefont {Varkentina}\ \emph {et~al.}(2023)\citenamefont
  {Varkentina}, \citenamefont {Auad}, \citenamefont {Woo}, \citenamefont
  {Castioni}, \citenamefont {Blazit}, \citenamefont {Tencé}, \citenamefont
  {Chang}, \citenamefont {Chen}, \citenamefont {Watanabe}, \citenamefont
  {Taniguchi}, \citenamefont {Kociak},\ and\ \citenamefont
  {Tizei}}]{varkentina2023excitations}%
  \BibitemOpen
  \bibfield  {author} {\bibinfo {author} {\bibfnamefont {N.}~\bibnamefont
  {Varkentina}}, \bibinfo {author} {\bibfnamefont {Y.}~\bibnamefont {Auad}},
  \bibinfo {author} {\bibfnamefont {S.~Y.}\ \bibnamefont {Woo}}, \bibinfo
  {author} {\bibfnamefont {F.}~\bibnamefont {Castioni}}, \bibinfo {author}
  {\bibfnamefont {J.-D.}\ \bibnamefont {Blazit}}, \bibinfo {author}
  {\bibfnamefont {M.}~\bibnamefont {Tencé}}, \bibinfo {author} {\bibfnamefont
  {H.-C.}\ \bibnamefont {Chang}}, \bibinfo {author} {\bibfnamefont
  {J.}~\bibnamefont {Chen}}, \bibinfo {author} {\bibfnamefont {K.}~\bibnamefont
  {Watanabe}}, \bibinfo {author} {\bibfnamefont {T.}~\bibnamefont {Taniguchi}},
  \bibinfo {author} {\bibfnamefont {M.}~\bibnamefont {Kociak}},\ and\ \bibinfo
  {author} {\bibfnamefont {L.~H.~G.}\ \bibnamefont {Tizei}},\ }\bibfield
  {title} {\bibinfo {title} {Excitation's lifetime extracted from
  electron-photon (eels-cl) nanosecond-scale temporal coincidences},\
  }\bibfield  {journal} {\bibinfo  {journal} {arXiv preprint arXiv:2306.15372}\
  }\href {https://doi.org/10.48550/arXiv.2306.15372}
  {10.48550/arXiv.2306.15372} (\bibinfo {year} {2023})\BibitemShut {NoStop}%
\bibitem [{\citenamefont {de~Abajo}\ and\ \citenamefont
  {Kociak}(2008)}]{de2008electron}%
  \BibitemOpen
  \bibfield  {author} {\bibinfo {author} {\bibfnamefont {F.~G.}\ \bibnamefont
  {de~Abajo}}\ and\ \bibinfo {author} {\bibfnamefont {M.}~\bibnamefont
  {Kociak}},\ }\bibfield  {title} {\bibinfo {title} {Electron energy-gain
  spectroscopy},\ }\href {https://doi.org/10.1088/1367-2630/10/7/073035}
  {\bibfield  {journal} {\bibinfo  {journal} {New Journal of Physics}\ }\textbf
  {\bibinfo {volume} {10}},\ \bibinfo {pages} {073035} (\bibinfo {year}
  {2008})}\BibitemShut {NoStop}%
\bibitem [{\citenamefont {Das}\ \emph {et~al.}(2019)\citenamefont {Das},
  \citenamefont {Blazit}, \citenamefont {Tenc{\'e}}, \citenamefont {Zagonel},
  \citenamefont {Auad}, \citenamefont {Losquin}, \citenamefont {Colliex},
  \citenamefont {St{\'e}phan} \emph {et~al.}}]{das2019stimulated}%
  \BibitemOpen
  \bibfield  {author} {\bibinfo {author} {\bibfnamefont {P.}~\bibnamefont
  {Das}}, \bibinfo {author} {\bibfnamefont {J.}~\bibnamefont {Blazit}},
  \bibinfo {author} {\bibfnamefont {M.}~\bibnamefont {Tenc{\'e}}}, \bibinfo
  {author} {\bibfnamefont {L.}~\bibnamefont {Zagonel}}, \bibinfo {author}
  {\bibfnamefont {Y.}~\bibnamefont {Auad}}, \bibinfo {author} {\bibfnamefont
  {A.}~\bibnamefont {Losquin}}, \bibinfo {author} {\bibfnamefont
  {C.}~\bibnamefont {Colliex}}, \bibinfo {author} {\bibfnamefont
  {O.}~\bibnamefont {St{\'e}phan}}, \emph {et~al.},\ }\bibfield  {title}
  {\bibinfo {title} {Stimulated electron energy loss and gain in an electron
  microscope without a pulsed electron gun},\ }\href
  {https://doi.org/10.1016/j.ultramic.2018.12.011} {\bibfield  {journal}
  {\bibinfo  {journal} {Ultramicroscopy}\ }\textbf {\bibinfo {volume} {203}},\
  \bibinfo {pages} {44} (\bibinfo {year} {2019})}\BibitemShut {NoStop}%
\bibitem [{\citenamefont {Auad}\ \emph {et~al.}(2023)\citenamefont {Auad},
  \citenamefont {Dias}, \citenamefont {Tenc{\'e}}, \citenamefont {Blazit},
  \citenamefont {Li}, \citenamefont {Zagonel}, \citenamefont {St{\'e}phan},
  \citenamefont {Tizei}, \citenamefont {Garc{\'\i}a~de Abajo},\ and\
  \citenamefont {Kociak}}]{auad2023muev}%
  \BibitemOpen
  \bibfield  {author} {\bibinfo {author} {\bibfnamefont {Y.}~\bibnamefont
  {Auad}}, \bibinfo {author} {\bibfnamefont {E.~J.}\ \bibnamefont {Dias}},
  \bibinfo {author} {\bibfnamefont {M.}~\bibnamefont {Tenc{\'e}}}, \bibinfo
  {author} {\bibfnamefont {J.-D.}\ \bibnamefont {Blazit}}, \bibinfo {author}
  {\bibfnamefont {X.}~\bibnamefont {Li}}, \bibinfo {author} {\bibfnamefont
  {L.~F.}\ \bibnamefont {Zagonel}}, \bibinfo {author} {\bibfnamefont
  {O.}~\bibnamefont {St{\'e}phan}}, \bibinfo {author} {\bibfnamefont {L.~H.}\
  \bibnamefont {Tizei}}, \bibinfo {author} {\bibfnamefont {F.~J.}\ \bibnamefont
  {Garc{\'\i}a~de Abajo}},\ and\ \bibinfo {author} {\bibfnamefont
  {M.}~\bibnamefont {Kociak}},\ }\bibfield  {title} {\bibinfo {title} {$\mu$ev
  electron spectromicroscopy using free-space light},\ }\href
  {https://doi.org/10.1038/s41467-023-39979-0} {\bibfield  {journal} {\bibinfo
  {journal} {Nature Communications}\ }\textbf {\bibinfo {volume} {14}},\
  \bibinfo {pages} {4442} (\bibinfo {year} {2023})}\BibitemShut {NoStop}%
\bibitem [{\citenamefont {Jakubek}(2011)}]{jakubek2011precise}%
  \BibitemOpen
  \bibfield  {author} {\bibinfo {author} {\bibfnamefont {J.}~\bibnamefont
  {Jakubek}},\ }\bibfield  {title} {\bibinfo {title} {Precise energy
  calibration of pixel detector working in time-over-threshold mode},\ }\href
  {https://doi.org/10.1016/j.nima.2010.06.183} {\bibfield  {journal} {\bibinfo
  {journal} {Nuclear Instruments and Methods in Physics Research Section A:
  Accelerators, Spectrometers, Detectors and Associated Equipment}\ }\textbf
  {\bibinfo {volume} {633}},\ \bibinfo {pages} {S262} (\bibinfo {year}
  {2011})}\BibitemShut {NoStop}%
\bibitem [{\citenamefont {Turecek}\ \emph {et~al.}(2016)\citenamefont
  {Turecek}, \citenamefont {Jakubek},\ and\ \citenamefont
  {Soukup}}]{turecek2016usb}%
  \BibitemOpen
  \bibfield  {author} {\bibinfo {author} {\bibfnamefont {D.}~\bibnamefont
  {Turecek}}, \bibinfo {author} {\bibfnamefont {J.}~\bibnamefont {Jakubek}},\
  and\ \bibinfo {author} {\bibfnamefont {P.}~\bibnamefont {Soukup}},\
  }\bibfield  {title} {\bibinfo {title} {Usb 3.0 readout and time-walk
  correction method for timepix3 detector},\ }\href
  {https://doi.org/10.1088/1748-0221/11/12/C12065} {\bibfield  {journal}
  {\bibinfo  {journal} {Journal of Instrumentation}\ }\textbf {\bibinfo
  {volume} {11}}\bibinfo  {number} { (12)},\ \bibinfo {pages}
  {C12065}}\BibitemShut {NoStop}%
\bibitem [{\citenamefont {Bergmann}\ \emph {et~al.}(2017)\citenamefont
  {Bergmann}, \citenamefont {Pichotka}, \citenamefont {Pospisil}, \citenamefont
  {Vycpalek}, \citenamefont {Burian}, \citenamefont {Broulim},\ and\
  \citenamefont {Jakubek}}]{bergmann20173d}%
  \BibitemOpen
\bibfield  {number} {  }\bibfield  {author} {\bibinfo {author} {\bibfnamefont
  {B.}~\bibnamefont {Bergmann}}, \bibinfo {author} {\bibfnamefont
  {M.}~\bibnamefont {Pichotka}}, \bibinfo {author} {\bibfnamefont
  {S.}~\bibnamefont {Pospisil}}, \bibinfo {author} {\bibfnamefont
  {J.}~\bibnamefont {Vycpalek}}, \bibinfo {author} {\bibfnamefont
  {P.}~\bibnamefont {Burian}}, \bibinfo {author} {\bibfnamefont
  {P.}~\bibnamefont {Broulim}},\ and\ \bibinfo {author} {\bibfnamefont
  {J.}~\bibnamefont {Jakubek}},\ }\bibfield  {title} {\bibinfo {title} {{3D}
  track reconstruction capability of a silicon hybrid active pixel detector},\
  }\href {https://doi.org/10.1140/epjc/s10052-017-4993-4} {\bibfield  {journal}
  {\bibinfo  {journal} {The European Physical Journal C}\ }\textbf {\bibinfo
  {volume} {77}},\ \bibinfo {pages} {1} (\bibinfo {year} {2017})}\BibitemShut
  {NoStop}%
\bibitem [{\citenamefont {Pitters}\ \emph {et~al.}(2019)\citenamefont
  {Pitters}, \citenamefont {Tehrani}, \citenamefont {Dannheim}, \citenamefont
  {Fiergolski}, \citenamefont {Hynds}, \citenamefont {Klempt}, \citenamefont
  {Llopart}, \citenamefont {Munker}, \citenamefont {N{\"u}rnberg},
  \citenamefont {Spannagel} \emph {et~al.}}]{pitters2019time}%
  \BibitemOpen
  \bibfield  {author} {\bibinfo {author} {\bibfnamefont {F.}~\bibnamefont
  {Pitters}}, \bibinfo {author} {\bibfnamefont {N.~A.}\ \bibnamefont
  {Tehrani}}, \bibinfo {author} {\bibfnamefont {D.}~\bibnamefont {Dannheim}},
  \bibinfo {author} {\bibfnamefont {A.}~\bibnamefont {Fiergolski}}, \bibinfo
  {author} {\bibfnamefont {D.}~\bibnamefont {Hynds}}, \bibinfo {author}
  {\bibfnamefont {W.}~\bibnamefont {Klempt}}, \bibinfo {author} {\bibfnamefont
  {X.}~\bibnamefont {Llopart}}, \bibinfo {author} {\bibfnamefont
  {M.}~\bibnamefont {Munker}}, \bibinfo {author} {\bibfnamefont
  {A.}~\bibnamefont {N{\"u}rnberg}}, \bibinfo {author} {\bibfnamefont
  {S.}~\bibnamefont {Spannagel}}, \emph {et~al.},\ }\bibfield  {title}
  {\bibinfo {title} {Time resolution studies of {T}imepix3 assemblies with thin
  silicon pixel sensors},\ }\href
  {https://doi.org/10.1088/1748-0221/14/05/P05022} {\bibfield  {journal}
  {\bibinfo  {journal} {Journal of Instrumentation}\ }\textbf {\bibinfo
  {volume} {14}}\bibinfo  {number} { (05)},\ \bibinfo {pages}
  {P05022}}\BibitemShut {NoStop}%
\bibitem [{\citenamefont {Wen}\ \emph {et~al.}(2022)\citenamefont {Wen},
  \citenamefont {Zheng}, \citenamefont {Gao}, \citenamefont {Zeng},
  \citenamefont {Zhang}, \citenamefont {Yu}, \citenamefont {Wu}, \citenamefont
  {Cang}, \citenamefont {Ma},\ and\ \citenamefont
  {Zhao}}]{wen2022optimization}%
  \BibitemOpen
\bibfield  {number} {  }\bibfield  {author} {\bibinfo {author} {\bibfnamefont
  {J.}~\bibnamefont {Wen}}, \bibinfo {author} {\bibfnamefont {X.}~\bibnamefont
  {Zheng}}, \bibinfo {author} {\bibfnamefont {H.}~\bibnamefont {Gao}}, \bibinfo
  {author} {\bibfnamefont {M.}~\bibnamefont {Zeng}}, \bibinfo {author}
  {\bibfnamefont {Y.}~\bibnamefont {Zhang}}, \bibinfo {author} {\bibfnamefont
  {M.}~\bibnamefont {Yu}}, \bibinfo {author} {\bibfnamefont {Y.}~\bibnamefont
  {Wu}}, \bibinfo {author} {\bibfnamefont {J.}~\bibnamefont {Cang}}, \bibinfo
  {author} {\bibfnamefont {G.}~\bibnamefont {Ma}},\ and\ \bibinfo {author}
  {\bibfnamefont {Z.}~\bibnamefont {Zhao}},\ }\bibfield  {title} {\bibinfo
  {title} {Optimization of timepix3-based conventional compton camera using
  electron track algorithm},\ }\href
  {https://doi.org/https://doi.org/10.1016/j.nima.2021.165954} {\bibfield
  {journal} {\bibinfo  {journal} {Nuclear Instruments and Methods in Physics
  Research Section A: Accelerators, Spectrometers, Detectors and Associated
  Equipment}\ }\textbf {\bibinfo {volume} {1021}},\ \bibinfo {pages} {165954}
  (\bibinfo {year} {2022})}\BibitemShut {NoStop}%
\bibitem [{\citenamefont {Meuret}\ \emph {et~al.}(2015)\citenamefont {Meuret},
  \citenamefont {Tizei}, \citenamefont {Cazimajou}, \citenamefont
  {Bourrellier}, \citenamefont {Chang}, \citenamefont {Treussart},\ and\
  \citenamefont {Kociak}}]{meuret2015photon}%
  \BibitemOpen
  \bibfield  {author} {\bibinfo {author} {\bibfnamefont {S.}~\bibnamefont
  {Meuret}}, \bibinfo {author} {\bibfnamefont {L.}~\bibnamefont {Tizei}},
  \bibinfo {author} {\bibfnamefont {T.}~\bibnamefont {Cazimajou}}, \bibinfo
  {author} {\bibfnamefont {R.}~\bibnamefont {Bourrellier}}, \bibinfo {author}
  {\bibfnamefont {H.}~\bibnamefont {Chang}}, \bibinfo {author} {\bibfnamefont
  {F.}~\bibnamefont {Treussart}},\ and\ \bibinfo {author} {\bibfnamefont
  {M.}~\bibnamefont {Kociak}},\ }\bibfield  {title} {\bibinfo {title} {Photon
  bunching in cathodoluminescence},\ }\href
  {https://doi.org/10.1103/PhysRevLett.114.197401} {\bibfield  {journal}
  {\bibinfo  {journal} {Physical review letters}\ }\textbf {\bibinfo {volume}
  {114}},\ \bibinfo {pages} {197401} (\bibinfo {year} {2015})}\BibitemShut
  {NoStop}%
\bibitem [{\citenamefont {Auad}\ \emph
  {et~al.}(2022{\natexlab{b}})\citenamefont {Auad}, \citenamefont {Kociak},
  \citenamefont {Tizei}, \citenamefont {Blazit}, \citenamefont {Walls},
  \citenamefont {Stéphan}, \citenamefont {De~la Peña},\ and\ \citenamefont
  {Tencé}}]{auad2022tp3tools}%
  \BibitemOpen
  \bibfield  {author} {\bibinfo {author} {\bibfnamefont {Y.}~\bibnamefont
  {Auad}}, \bibinfo {author} {\bibfnamefont {M.}~\bibnamefont {Kociak}},
  \bibinfo {author} {\bibfnamefont {L.~H.~G.}\ \bibnamefont {Tizei}}, \bibinfo
  {author} {\bibfnamefont {J.-D.}\ \bibnamefont {Blazit}}, \bibinfo {author}
  {\bibfnamefont {M.}~\bibnamefont {Walls}}, \bibinfo {author} {\bibfnamefont
  {O.}~\bibnamefont {Stéphan}}, \bibinfo {author} {\bibfnamefont
  {F.}~\bibnamefont {De~la Peña}},\ and\ \bibinfo {author} {\bibfnamefont
  {M.}~\bibnamefont {Tencé}},\ }\href {https://doi.org/10.5281/zenodo.6346261}
  {\bibinfo {title} {Timestem/tp3\_tools: Release v1.0.0}} (\bibinfo {year}
  {2022}{\natexlab{b}})\BibitemShut {NoStop}%
\bibitem [{\citenamefont {Heijne}\ \emph {et~al.}(2013)\citenamefont {Heijne},
  \citenamefont {Sune}, \citenamefont {Campbell}, \citenamefont {Leroy},
  \citenamefont {Llopart}, \citenamefont {Martin}, \citenamefont {Pospisil},
  \citenamefont {Solc}, \citenamefont {Soueid}, \citenamefont {Suk} \emph
  {et~al.}}]{heijne2013measuring}%
  \BibitemOpen
  \bibfield  {author} {\bibinfo {author} {\bibfnamefont {E.~H.}\ \bibnamefont
  {Heijne}}, \bibinfo {author} {\bibfnamefont {R.~B.}\ \bibnamefont {Sune}},
  \bibinfo {author} {\bibfnamefont {M.}~\bibnamefont {Campbell}}, \bibinfo
  {author} {\bibfnamefont {C.}~\bibnamefont {Leroy}}, \bibinfo {author}
  {\bibfnamefont {X.}~\bibnamefont {Llopart}}, \bibinfo {author} {\bibfnamefont
  {J.-P.}\ \bibnamefont {Martin}}, \bibinfo {author} {\bibfnamefont
  {S.}~\bibnamefont {Pospisil}}, \bibinfo {author} {\bibfnamefont
  {J.}~\bibnamefont {Solc}}, \bibinfo {author} {\bibfnamefont {P.}~\bibnamefont
  {Soueid}}, \bibinfo {author} {\bibfnamefont {M.}~\bibnamefont {Suk}}, \emph
  {et~al.},\ }\bibfield  {title} {\bibinfo {title} {Measuring radiation
  environment in lhc or anywhere else, on your computer screen with medipix},\
  }\href {https://doi.org/10.1016/j.nima.2012.05.023} {\bibfield  {journal}
  {\bibinfo  {journal} {Nuclear Instruments and Methods in Physics Research
  Section A: Accelerators, Spectrometers, Detectors and Associated Equipment}\
  }\textbf {\bibinfo {volume} {699}},\ \bibinfo {pages} {198} (\bibinfo {year}
  {2013})}\BibitemShut {NoStop}%
\bibitem [{\citenamefont {Granja}\ \emph {et~al.}(2018)\citenamefont {Granja},
  \citenamefont {Kudela}, \citenamefont {Jakubek}, \citenamefont {Krist},
  \citenamefont {Chvatil}, \citenamefont {Stursa},\ and\ \citenamefont
  {Polansky}}]{granja2018directional}%
  \BibitemOpen
  \bibfield  {author} {\bibinfo {author} {\bibfnamefont {C.}~\bibnamefont
  {Granja}}, \bibinfo {author} {\bibfnamefont {K.}~\bibnamefont {Kudela}},
  \bibinfo {author} {\bibfnamefont {J.}~\bibnamefont {Jakubek}}, \bibinfo
  {author} {\bibfnamefont {P.}~\bibnamefont {Krist}}, \bibinfo {author}
  {\bibfnamefont {D.}~\bibnamefont {Chvatil}}, \bibinfo {author} {\bibfnamefont
  {J.}~\bibnamefont {Stursa}},\ and\ \bibinfo {author} {\bibfnamefont
  {S.}~\bibnamefont {Polansky}},\ }\bibfield  {title} {\bibinfo {title}
  {Directional detection of charged particles and cosmic rays with the
  miniaturized radiation camera minipix timepix},\ }\href
  {https://doi.org/10.1016/j.nima.2018.09.140} {\bibfield  {journal} {\bibinfo
  {journal} {Nuclear Instruments and Methods in Physics Research Section A:
  Accelerators, Spectrometers, Detectors and Associated Equipment}\ }\textbf
  {\bibinfo {volume} {911}},\ \bibinfo {pages} {142} (\bibinfo {year}
  {2018})}\BibitemShut {NoStop}%
\bibitem [{\citenamefont {Hovington}\ \emph {et~al.}(1997)\citenamefont
  {Hovington}, \citenamefont {Drouin},\ and\ \citenamefont
  {Gauvin}}]{hovington1997casino}%
  \BibitemOpen
  \bibfield  {author} {\bibinfo {author} {\bibfnamefont {P.}~\bibnamefont
  {Hovington}}, \bibinfo {author} {\bibfnamefont {D.}~\bibnamefont {Drouin}},\
  and\ \bibinfo {author} {\bibfnamefont {R.}~\bibnamefont {Gauvin}},\
  }\bibfield  {title} {\bibinfo {title} {Casino: A new monte carlo code in c
  language for electron beam interaction—part i: Description of the
  program},\ }\href {https://doi.org/10.1002/sca.4950190101} {\bibfield
  {journal} {\bibinfo  {journal} {Scanning}\ }\textbf {\bibinfo {volume}
  {19}},\ \bibinfo {pages} {1} (\bibinfo {year} {1997})}\BibitemShut {NoStop}%
\bibitem [{\citenamefont {Agostinelli}\ \emph {et~al.}(2003)\citenamefont
  {Agostinelli}, \citenamefont {Allison}, \citenamefont {Amako}, \citenamefont
  {Apostolakis}, \citenamefont {Araujo}, \citenamefont {Arce}, \citenamefont
  {Asai}, \citenamefont {Axen}, \citenamefont {Banerjee}, \citenamefont
  {Barrand} \emph {et~al.}}]{agostinelli2003geant4}%
  \BibitemOpen
  \bibfield  {author} {\bibinfo {author} {\bibfnamefont {S.}~\bibnamefont
  {Agostinelli}}, \bibinfo {author} {\bibfnamefont {J.}~\bibnamefont
  {Allison}}, \bibinfo {author} {\bibfnamefont {K.~a.}\ \bibnamefont {Amako}},
  \bibinfo {author} {\bibfnamefont {J.}~\bibnamefont {Apostolakis}}, \bibinfo
  {author} {\bibfnamefont {H.}~\bibnamefont {Araujo}}, \bibinfo {author}
  {\bibfnamefont {P.}~\bibnamefont {Arce}}, \bibinfo {author} {\bibfnamefont
  {M.}~\bibnamefont {Asai}}, \bibinfo {author} {\bibfnamefont {D.}~\bibnamefont
  {Axen}}, \bibinfo {author} {\bibfnamefont {S.}~\bibnamefont {Banerjee}},
  \bibinfo {author} {\bibfnamefont {G.}~\bibnamefont {Barrand}}, \emph
  {et~al.},\ }\bibfield  {title} {\bibinfo {title} {Geant4—a simulation
  toolkit},\ }\href {https://doi.org/10.1016/S0168-9002(03)01368-8} {\bibfield
  {journal} {\bibinfo  {journal} {Nuclear instruments and methods in physics
  research section A: Accelerators, Spectrometers, Detectors and Associated
  Equipment}\ }\textbf {\bibinfo {volume} {506}},\ \bibinfo {pages} {250}
  (\bibinfo {year} {2003})}\BibitemShut {NoStop}%
\bibitem [{\citenamefont {Sch{\"u}bel}\ \emph {et~al.}(2014)\citenamefont
  {Sch{\"u}bel}, \citenamefont {Krapohl}, \citenamefont {Fr{\"o}jdh},
  \citenamefont {Fr{\"o}jdh},\ and\ \citenamefont
  {Thungstr{\"o}m}}]{schubel2014geant4}%
  \BibitemOpen
  \bibfield  {author} {\bibinfo {author} {\bibfnamefont {A.}~\bibnamefont
  {Sch{\"u}bel}}, \bibinfo {author} {\bibfnamefont {D.}~\bibnamefont
  {Krapohl}}, \bibinfo {author} {\bibfnamefont {E.}~\bibnamefont {Fr{\"o}jdh}},
  \bibinfo {author} {\bibfnamefont {C.}~\bibnamefont {Fr{\"o}jdh}},\ and\
  \bibinfo {author} {\bibfnamefont {G.}~\bibnamefont {Thungstr{\"o}m}},\
  }\bibfield  {title} {\bibinfo {title} {A geant4 based framework for pixel
  detector simulation},\ }\href {https://doi.org/10.1088/1748-0221/9/12/C12018}
  {\bibfield  {journal} {\bibinfo  {journal} {Journal of Instrumentation}\
  }\textbf {\bibinfo {volume} {9}}\bibinfo  {number} { (12)},\ \bibinfo {pages}
  {C12018}}\BibitemShut {NoStop}%
\bibitem [{\citenamefont {Feist}\ \emph {et~al.}(2017)\citenamefont {Feist},
  \citenamefont {Bach}, \citenamefont {da~Silva}, \citenamefont {Danz},
  \citenamefont {M{\"o}ller}, \citenamefont {Priebe}, \citenamefont
  {Domr{\"o}se}, \citenamefont {Gatzmann}, \citenamefont {Rost}, \citenamefont
  {Schauss} \emph {et~al.}}]{feist2017ultrafast}%
  \BibitemOpen
\bibfield  {number} {  }\bibfield  {author} {\bibinfo {author} {\bibfnamefont
  {A.}~\bibnamefont {Feist}}, \bibinfo {author} {\bibfnamefont
  {N.}~\bibnamefont {Bach}}, \bibinfo {author} {\bibfnamefont {N.~R.}\
  \bibnamefont {da~Silva}}, \bibinfo {author} {\bibfnamefont {T.}~\bibnamefont
  {Danz}}, \bibinfo {author} {\bibfnamefont {M.}~\bibnamefont {M{\"o}ller}},
  \bibinfo {author} {\bibfnamefont {K.~E.}\ \bibnamefont {Priebe}}, \bibinfo
  {author} {\bibfnamefont {T.}~\bibnamefont {Domr{\"o}se}}, \bibinfo {author}
  {\bibfnamefont {J.~G.}\ \bibnamefont {Gatzmann}}, \bibinfo {author}
  {\bibfnamefont {S.}~\bibnamefont {Rost}}, \bibinfo {author} {\bibfnamefont
  {J.}~\bibnamefont {Schauss}}, \emph {et~al.},\ }\bibfield  {title} {\bibinfo
  {title} {Ultrafast transmission electron microscopy using a laser-driven
  field emitter: Femtosecond resolution with a high coherence electron beam},\
  }\href {https://doi.org/10.1016/j.ultramic.2016.12.005} {\bibfield  {journal}
  {\bibinfo  {journal} {Ultramicroscopy}\ }\textbf {\bibinfo {volume} {176}},\
  \bibinfo {pages} {63} (\bibinfo {year} {2017})}\BibitemShut {NoStop}%
\bibitem [{\citenamefont {Houdellier}\ \emph {et~al.}(2018)\citenamefont
  {Houdellier}, \citenamefont {Caruso}, \citenamefont {Weber}, \citenamefont
  {Kociak},\ and\ \citenamefont {Arbouet}}]{houdellier2018development}%
  \BibitemOpen
  \bibfield  {author} {\bibinfo {author} {\bibfnamefont {F.}~\bibnamefont
  {Houdellier}}, \bibinfo {author} {\bibfnamefont {G.~M.}\ \bibnamefont
  {Caruso}}, \bibinfo {author} {\bibfnamefont {S.}~\bibnamefont {Weber}},
  \bibinfo {author} {\bibfnamefont {M.}~\bibnamefont {Kociak}},\ and\ \bibinfo
  {author} {\bibfnamefont {A.}~\bibnamefont {Arbouet}},\ }\bibfield  {title}
  {\bibinfo {title} {Development of a high brightness ultrafast transmission
  electron microscope based on a laser-driven cold field emission source},\
  }\href {https://doi.org/10.1016/j.ultramic.2017.12.015} {\bibfield  {journal}
  {\bibinfo  {journal} {Ultramicroscopy}\ }\textbf {\bibinfo {volume} {186}},\
  \bibinfo {pages} {128} (\bibinfo {year} {2018})}\BibitemShut {NoStop}%
\bibitem [{\citenamefont {Campbell}\ \emph {et~al.}(2016)\citenamefont
  {Campbell}, \citenamefont {Alozy}, \citenamefont {Ballabriga}, \citenamefont
  {Frojdh}, \citenamefont {Heijne}, \citenamefont {Llopart}, \citenamefont
  {Poikela}, \citenamefont {Tlustos}, \citenamefont {Valerio},\ and\
  \citenamefont {Wong}}]{campbell2016towards}%
  \BibitemOpen
  \bibfield  {author} {\bibinfo {author} {\bibfnamefont {M.}~\bibnamefont
  {Campbell}}, \bibinfo {author} {\bibfnamefont {J.}~\bibnamefont {Alozy}},
  \bibinfo {author} {\bibfnamefont {R.}~\bibnamefont {Ballabriga}}, \bibinfo
  {author} {\bibfnamefont {E.}~\bibnamefont {Frojdh}}, \bibinfo {author}
  {\bibfnamefont {E.}~\bibnamefont {Heijne}}, \bibinfo {author} {\bibfnamefont
  {X.}~\bibnamefont {Llopart}}, \bibinfo {author} {\bibfnamefont
  {T.}~\bibnamefont {Poikela}}, \bibinfo {author} {\bibfnamefont
  {L.}~\bibnamefont {Tlustos}}, \bibinfo {author} {\bibfnamefont
  {P.}~\bibnamefont {Valerio}},\ and\ \bibinfo {author} {\bibfnamefont
  {W.}~\bibnamefont {Wong}},\ }\bibfield  {title} {\bibinfo {title} {Towards a
  new generation of pixel detector readout chips},\ }\href
  {https://doi.org/10.1088/1748-0221/11/01/C01007} {\bibfield  {journal}
  {\bibinfo  {journal} {Journal of Instrumentation}\ }\textbf {\bibinfo
  {volume} {11}}\bibinfo  {number} { (01)},\ \bibinfo {pages}
  {C01007}}\BibitemShut {NoStop}%
\bibitem [{\citenamefont {Llopart}\ \emph {et~al.}(2022)\citenamefont
  {Llopart}, \citenamefont {Alozy}, \citenamefont {Ballabriga}, \citenamefont
  {Campbell}, \citenamefont {Casanova}, \citenamefont {Gromov}, \citenamefont
  {Heijne}, \citenamefont {Poikela}, \citenamefont {Santin}, \citenamefont
  {Sriskaran} \emph {et~al.}}]{llopart2022timepix4}%
  \BibitemOpen
\bibfield  {number} {  }\bibfield  {author} {\bibinfo {author} {\bibfnamefont
  {X.}~\bibnamefont {Llopart}}, \bibinfo {author} {\bibfnamefont
  {J.}~\bibnamefont {Alozy}}, \bibinfo {author} {\bibfnamefont
  {R.}~\bibnamefont {Ballabriga}}, \bibinfo {author} {\bibfnamefont
  {M.}~\bibnamefont {Campbell}}, \bibinfo {author} {\bibfnamefont
  {R.}~\bibnamefont {Casanova}}, \bibinfo {author} {\bibfnamefont
  {V.}~\bibnamefont {Gromov}}, \bibinfo {author} {\bibfnamefont
  {E.}~\bibnamefont {Heijne}}, \bibinfo {author} {\bibfnamefont
  {T.}~\bibnamefont {Poikela}}, \bibinfo {author} {\bibfnamefont
  {E.}~\bibnamefont {Santin}}, \bibinfo {author} {\bibfnamefont
  {V.}~\bibnamefont {Sriskaran}}, \emph {et~al.},\ }\bibfield  {title}
  {\bibinfo {title} {Timepix4, a large area pixel detector readout chip which
  can be tiled on 4 sides providing sub-200 ps timestamp binning},\ }\href
  {https://doi.org/10.1088/1748-0221/17/01/C01044} {\bibfield  {journal}
  {\bibinfo  {journal} {Journal of Instrumentation}\ }\textbf {\bibinfo
  {volume} {17}}\bibinfo  {number} { (01)},\ \bibinfo {pages}
  {C01044}}\BibitemShut {NoStop}%
\bibitem [{\citenamefont {Heijhoff}\ \emph {et~al.}(2022)\citenamefont
  {Heijhoff}, \citenamefont {Akiba}, \citenamefont {Ballabriga}, \citenamefont
  {van Beuzekom}, \citenamefont {Campbell}, \citenamefont {Colijn},
  \citenamefont {Fransen}, \citenamefont {Geertsema}, \citenamefont {Gromov},\
  and\ \citenamefont {Cudie}}]{heijhoff2022timing}%
  \BibitemOpen
\bibfield  {number} {  }\bibfield  {author} {\bibinfo {author} {\bibfnamefont
  {K.}~\bibnamefont {Heijhoff}}, \bibinfo {author} {\bibfnamefont
  {K.}~\bibnamefont {Akiba}}, \bibinfo {author} {\bibfnamefont
  {R.}~\bibnamefont {Ballabriga}}, \bibinfo {author} {\bibfnamefont
  {M.}~\bibnamefont {van Beuzekom}}, \bibinfo {author} {\bibfnamefont
  {M.}~\bibnamefont {Campbell}}, \bibinfo {author} {\bibfnamefont
  {A.}~\bibnamefont {Colijn}}, \bibinfo {author} {\bibfnamefont
  {M.}~\bibnamefont {Fransen}}, \bibinfo {author} {\bibfnamefont
  {R.}~\bibnamefont {Geertsema}}, \bibinfo {author} {\bibfnamefont
  {V.}~\bibnamefont {Gromov}},\ and\ \bibinfo {author} {\bibfnamefont {X.~L.}\
  \bibnamefont {Cudie}},\ }\bibfield  {title} {\bibinfo {title} {Timing
  performance of the timepix4 front-end},\ }\href
  {https://doi.org/10.1088/1748-0221/17/07/P07006} {\bibfield  {journal}
  {\bibinfo  {journal} {Journal of Instrumentation}\ }\textbf {\bibinfo
  {volume} {17}}\bibinfo  {number} { (07)},\ \bibinfo {pages}
  {P07006}}\BibitemShut {NoStop}%
\end{thebibliography}%

\setcounter{figure}{0}
\setcounter{equation}{0}
\makeatletter 
\renewcommand{\thefigure}{S\@arabic\c@figure}
\makeatother
\onecolumngrid

\clearpage
\begin{large}
\textbf{Supplementary Material for Time calibration studies for the Timepix3 hybrid pixel detector in electron microscopy}
\end{large}

\section*{Energy calibration histograms}

\begin{figure}[h!]
    \centering
    \includegraphics[width=0.75\textwidth]{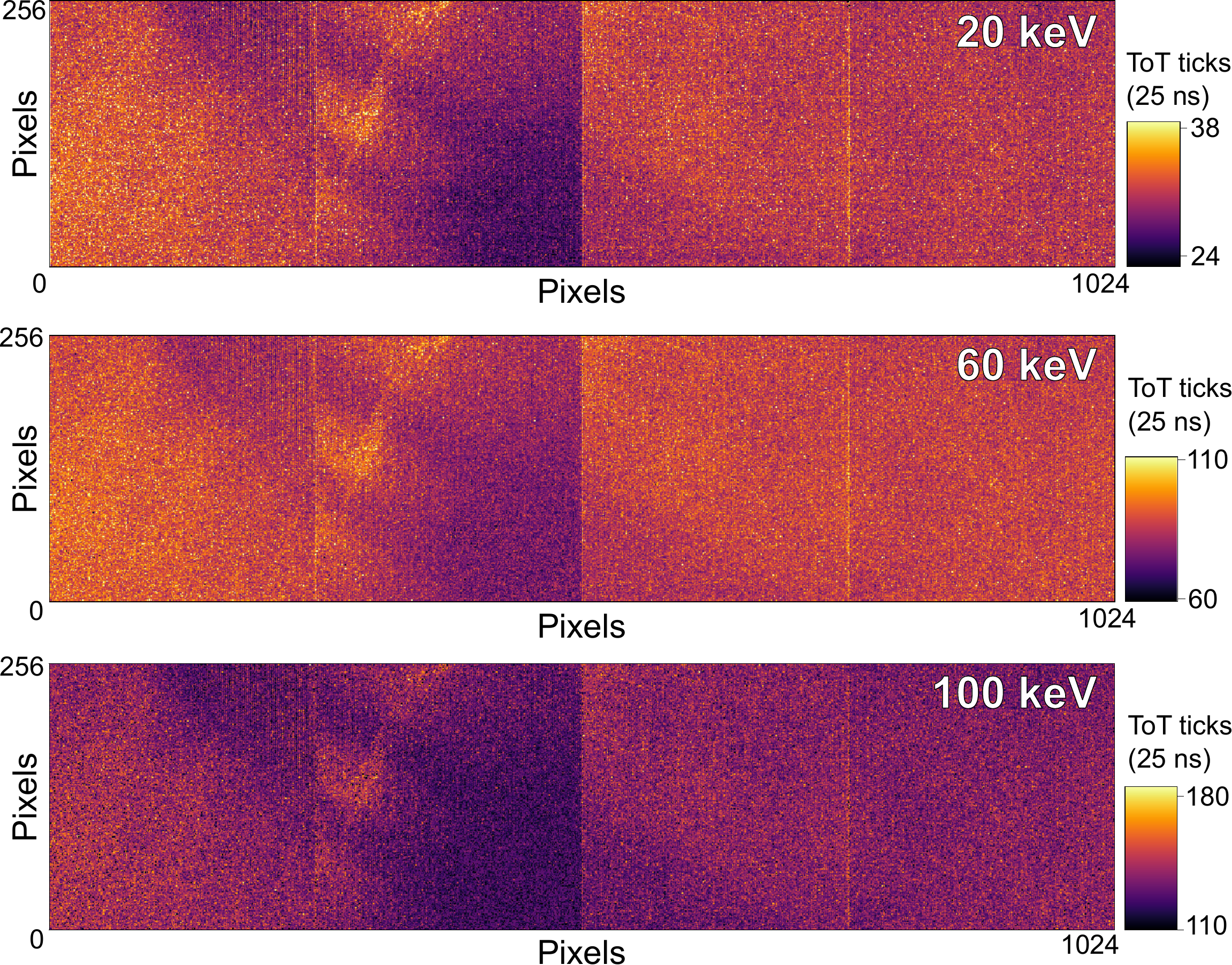}
    \caption{\textbf{Energy calibration histograms}. Data processing only considered clusters with a single hit, meaning that the electron energy was not shared with nearby pixels. For every pixel, the ToT distribution has been fitted with a Gaussian, and the center values are displayed here, in units of ToT clock ticks (25 ns) for 20 keV, 60 keV, and 100 keV electrons. A linear fit, per pixel, is used to correlate the energy deposited with the obtained ToT.}
    \label{FigureSup1}
\end{figure}

The energy deposited in the pixel is related to the time over the threshold (ToT). The detector can be calibrated in energy by considering only single-hit clusters in the data processing, meaning that the absorbed electron energy was not shared with nearby pixels. This has been done for 20 keV, 60 keV, 80 keV, and 100 keV electrons, in which the ToT histogram peak was fitted with a Gaussian, pixel by pixel. Figure \ref{FigureSup1} shows the Gaussian center for every pixel for three of the cited electron energies. With this, the coefficients d and e, as defined below, have been determined for each pixel.

\begin{equation}
    ToT(x, y, Energy) = d(x, y) \times Energy + e(x, y)
\end{equation}

As discussed in the main text, the dependence of the ToT on energy deviates from a linear model for low ToT values. As the minimum electron energy on the microscope is 20 keV, the lack of lower energy electrons undermines the study of this relation for low deposited energies. Fortunately, these low-energy hits have a poor temporal resolution due to the time-walk effect and are generally discarded from the dataset.

\clearpage

\section*{Time delay histograms}

\begin{figure}[h!]
    \centering
    \includegraphics[width=0.9\textwidth]{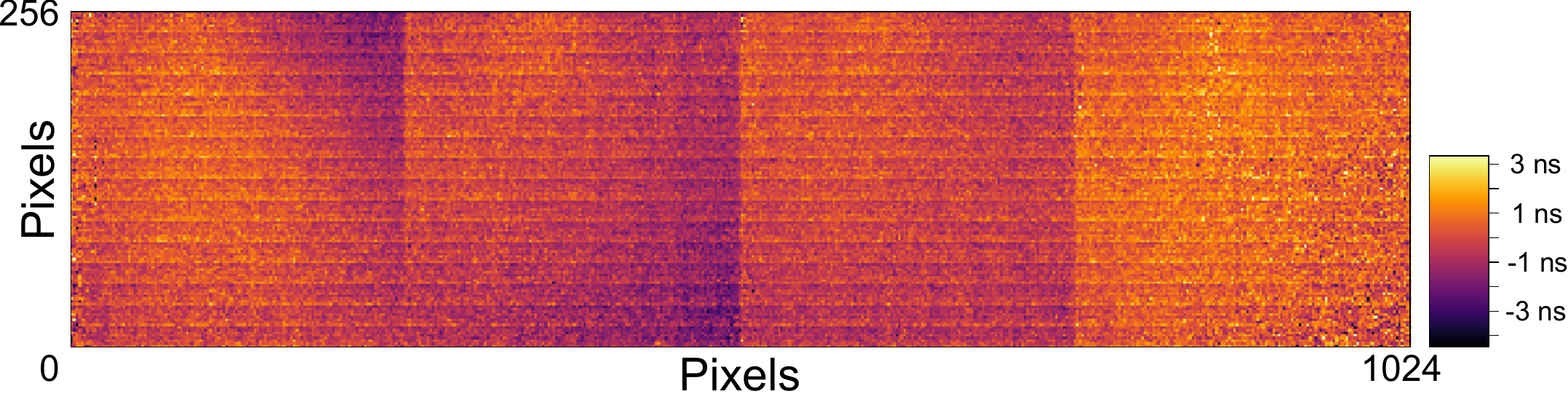}
    \caption{\textbf{Time delay histogram in the pixel array matrix}. Electron-photon coincidences have been obtained for each pixel and fitted with Gaussians. This histogram shows the center value of the Gaussians, which is used to correct the dataset time delay between pixels.}
    \label{FigureSup2A}
\end{figure}

\begin{figure}[h!]
    \centering
    \includegraphics[width=0.45\textwidth]{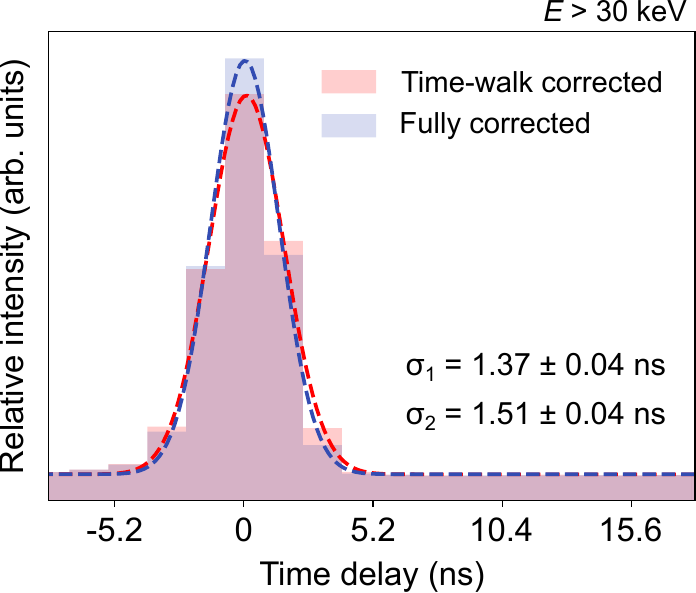}
    \caption{\textbf{Impact of the time delay calibration in the dataset}. Here we access the impact of the time delay calibration by comparing the partially corrected data (time-walk only) and the fully corrected data (time-walk + delay). The average temporal resolution is increased from $1.51 \pm 0.04$ ns to $1.37 \pm 0.04$ ns. Only energy hits with $E > 30$ keV have been considered for both curves.}
    \label{FigureSup2B}
\end{figure}

The time delay between pixels is obtained by Gaussian fitting the electron-photon coincidence histograms. The center of the Gaussians is shown in Figure \ref{FigureSup2A} and is compensated pixel-by-pixel. The obtained delays vary approximately from -4 ns to 3 ns around the average value, and the standard deviation of the entire histogram is 0.85 ns. In Figure \ref{FigureSup2B}, we show the impact of the time delay step calibration by comparing the time delay histograms after the time-walk correction and after the time-walk and the time delay (complete calibration). For this particular dataset, the overal temporal resolution increased from $1.51 \pm 0.04$ ns to $1.37 \pm 0.04$ ns. Similarly to the main text, only energy hits with $E > 30$ keV have been considered for both curves.

\clearpage

\section*{Lifetime values from the \textit{h}--BN sample}

The electron-photon coincidence histogram gives access to the decay's lifetime from the optical excitation. The curve's fall time is related to electrons arriving later in the ASIC, and it is mostly impacted by the time-walk effect. After correction, we obtain a value of $\tau_{fall} = 1.1 \pm 0.1$ ns, which is roughly our instrument response time. In the negative values, the photons are arriving later in the time-to-digital-converter, and thus the decay's lifetime can be accessed. This value is convoluted with the instrument response time, and here we have obtained a lifetime of $1.4 \pm 0.1$ ns, in which the sample's decay time measured by interferometry (120 ps resolution instrument) is $0.8 \pm 0.1$ ns. The comparison between $\tau_{fall}$ and $\tau_{rise}$ thus gives relevant and complementary information from our sample.

\begin{figure}[h!]
    \centering
    \includegraphics[width=0.45\textwidth]{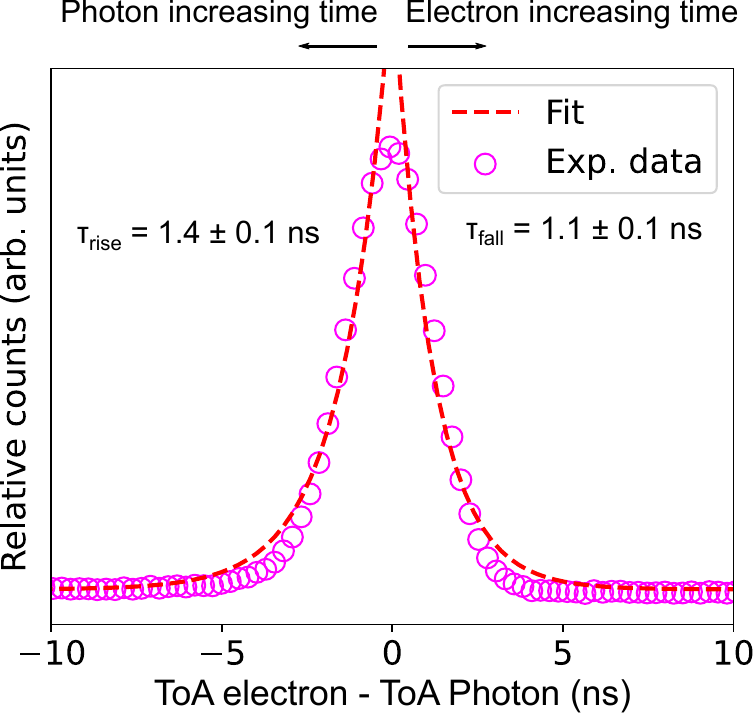}
    \caption{\textbf{Rise and fall times from electron-photon correlation histograms}. The rise and fall times of the correlation curves give physical information from our sample. Following the convention used throughout this work, in the direction of positive values (fall curve), the electrons arrive later (time-walk effect, for example). On the other side, in the rise time, photons arrive later, which gives the decay's lifetime of the optical excitation.}
    \label{FigureSup3}
\end{figure}

\clearpage

\end{document}